%% file: main.tex
\documentclass[conference,compsoc]{IEEEtran}
\usepackage{multirow}
\usepackage{booktabs}
\usepackage[dvipsnames]{xcolor}
\usepackage{soul}
\usepackage{hyperref}
\newcommand{\redcolorize}[2]{\colorbox{Salmon!#1!white}{\strut #2}}
\newcommand{\bluecolorize}[2]{\colorbox{Cyan!#1!white}{\strut #2}}
\newcommand{\reduline}[1]{\setulcolor{red}\ul{#1}}
\ifCLASSOPTIONcompsoc
  \usepackage[nocompress]{cite}
\else
  \usepackage{cite}
\fi
\ifCLASSINFOpdf
  \usepackage[pdftex]{graphicx}
\else
\fi
\usepackage{amsmath}
\usepackage{array}

\ifCLASSOPTIONcompsoc
 \usepackage[caption=false,font=footnotesize,labelfont=sf,textfont=sf]{subfig}
\else
 \usepackage[caption=false,font=footnotesize]{subfig}
\fi
\usepackage{url}

\begin{document}
%
\title{The Fire Thief Is Also the Keeper: \\Balancing Usability and Privacy in Prompts}

\author{
\IEEEauthorblockN{Zhili Shen, Zihang Xi, Ying He, Wei Tong, Jingyu Hua, Sheng Zhong}
\textit{Nanjing University}
\IEEEauthorblockA{\texttt{\{shenzhili, 201220212, dz1933010\}@smail.nju.edu.cn, weitong@outlook.com,}}
\IEEEauthorblockA{\texttt{\{huajingyu, zhongsheng\}@nju.edu.cn}}}


%


\maketitle

\begin{abstract}
The rapid adoption of online chatbots represents a significant advancement in artificial intelligence. However, this convenience brings considerable privacy concerns, as prompts can inadvertently contain sensitive information exposed to large language models (LLMs). Limited by high computational costs, reduced task usability, and excessive system modifications, previous works based on local deployment, embedding perturbation, and homomorphic encryption are inapplicable to online prompt-based LLM applications.

To address these issues, this paper introduces \texttt{\underline{Pro}mpt Privacy \underline{San}itizer} (i.e., \texttt{ProSan}), an end-to-end prompt privacy protection framework that can produce anonymized prompts with contextual privacy removed while maintaining task usability and human readability. It can also be seamlessly integrated into the online LLM service pipeline. To achieve high usability and dynamic anonymity, \texttt{ProSan} flexibly adjusts its protection targets and strength based on the importance of the words and the privacy leakage risk of the prompts. Additionally, \texttt{ProSan} is capable of adapting to diverse computational resource conditions, ensuring privacy protection even for mobile devices with limited computing power. Our experiments demonstrate that \texttt{ProSan} effectively removes private information across various tasks, including question answering, text summarization, and code generation, with minimal reduction in task performance.
\end{abstract}


%
\IEEEpeerreviewmaketitle

\section{Introduction}
The advent of online chatbots like ChatGPT is regarded as one of the most significant advances in the realm of general artificial intelligence\cite{qin2023chatgpt}. It also represents a major shift in the NLP application paradigm. Unlike fine-tuning\cite{devlin2019bert} or prompt-tuning\cite{lester2021power}, which requires data to adjust model parameters or prompts, large language models (LLMs) like GPT\cite{radford2019language} and LLaMA\cite{touvron2023llama} can be directly applied to downstream tasks without any such preparation. Users need only craft a prompt combining task instructions and relevant information, submit it to an LLM, and the model will complete the task in a zero-shot manner.

However, privacy leakage in prompts is one of the risks associated with LLM applications\cite{edemacu2024privacy}. To guide LLMs in completing tasks accurately, users need to provide relevant task information. The following types of information contained in prompts could pose privacy risks: \textbf{(1) Personal privacy information}, such as name, gender, or Social Security Number, may be disclosed when using healthcare systems based on LLMs; \textbf{(2) Corporate confidential information}, such as system log or financial statement, which could be revealed when employees use LLMs for tasks like coding or text summarization. Although service providers like OpenAI claim they will not disclose user data to third parties, they still collect user conversations as datasets for model training to enhance service quality\cite{openai2024privacy}. Existing attackers could extract or reconstruct user-private text through the model's API\cite{li2023multi}, parameters\cite{carlini2021extracting}, gradients\cite{gupta2022recovering}, and historical sessions\cite{chu2024conversation}. Thus, privacy leakage in prompts remains an obstacle to achieving secure general artificial intelligence.

\begin{figure}[t]
\centering
\includegraphics[width=0.45\textwidth]{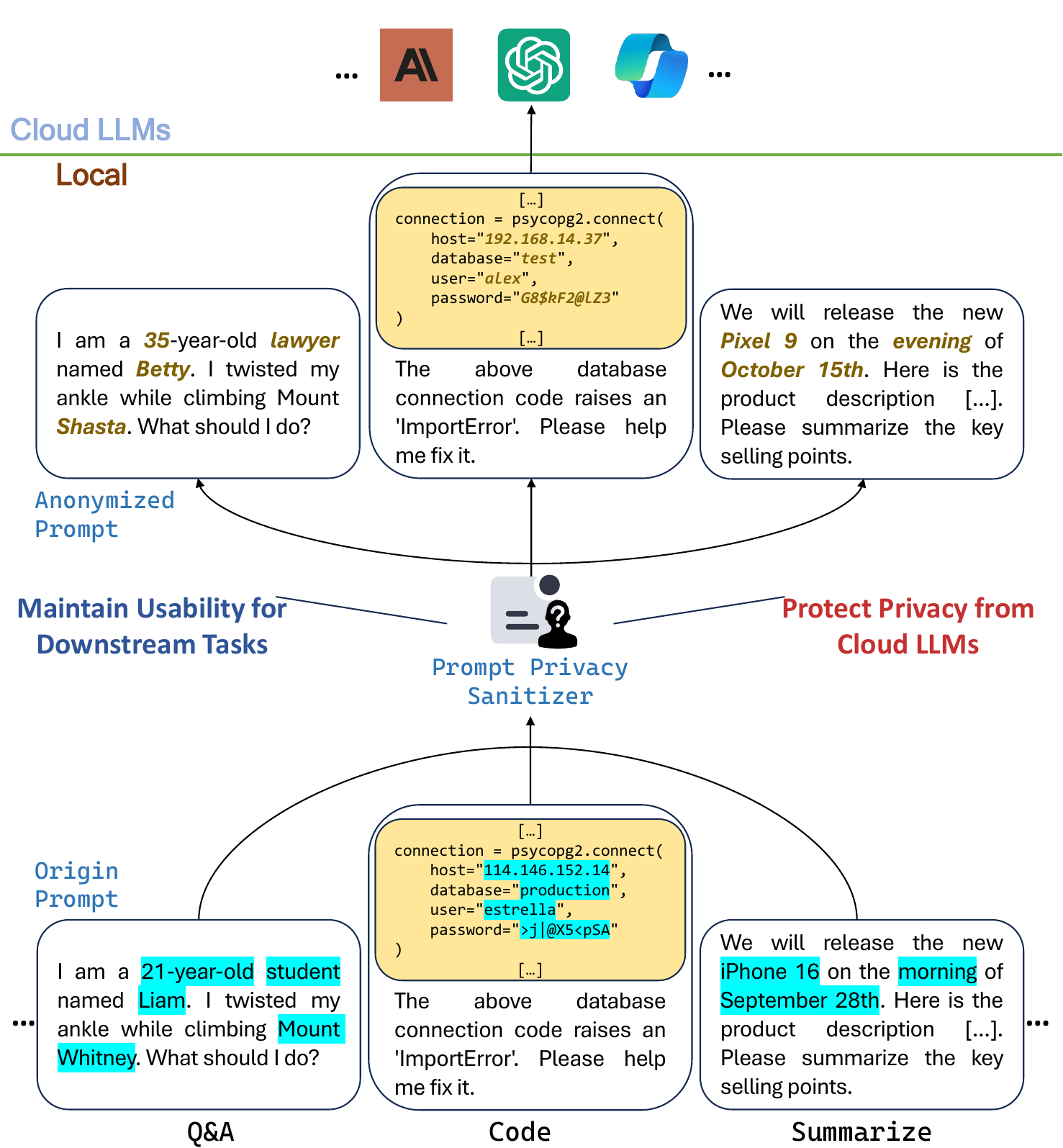}
\caption{\texttt{Prompt Privacy Sanitizer} illustration: \texttt{Prompt Privacy Sanitizer} measures the importance and privacy of words in the prompt and selectively replaces them. It can generate anonymized prompts for various tasks while maintaining high usability and dynamic anonymity.}
\label{fig:1_1}
\end{figure}

In the field of natural language processing (NLP), one research direction for text privacy protection is to prevent data from being explicitly exposed to service providers. This can be categorized into the following approaches: \textbf{(1) Local Deployment}: The most direct way to prevent privacy leakage is to run services offline, ensuring data never leaves the local environment. However, many commercial LLMs are closed-source. Previous work has executed privacy-related parts of tasks\cite{zhang2024cogenesis} or erased sensitive information\cite{kan2023protecting} in prompts by using locally hosted open-source LLMs. Nonetheless, even with model compression and acceleration technologies\cite{xu2023survey}, the high computational costs make it difficult for ordinary users to afford. \textbf{(2) Using Embeddings}\cite{huang2020texthide, zhou2022textfusion, zhou2023textmixer}: Another approach involves replacing the plaintext interface between the user and service providers with corresponding embeddings. However, this requires modifying other components in the NLP pipeline to adapt to this change. Additionally, because embeddings are not human-readable, users cannot verify if private information has truly been removed. Existing research indicates that attackers can reconstruct the original text or infer sensitive private attributes from text embeddings\cite{pan2020privacy}. \textbf{(3) Homomorphic Encryption}\cite{zhang2024privacyasst}: This technique allows service providers to process encrypted texts directly. However, it is challenging to apply to discrete texts, limiting its use cases. Furthermore, the encryption and decryption processes can impose additional computational burdens on the user side.

The aforementioned protection approaches all involve modifying the existing paradigm of LLM services, potentially introducing additional risks. Therefore, another direction of defensive measures is to directly delete private information in text at the user end. These measures can be categorized as follows: \textbf{(1) Differential Privacy}\cite{dwork2006differential}: Current work adds random noise into embeddings at word\cite{yue2021differential}, sentence\cite{du2023sanitizing}, and document\cite{meehan2022sentence} levels. By utilizing $k$-nearest neighbors\cite{feyisetan2020privacy}, autoencoders\cite{krishna2021adept}, or autoregressive language models\cite{mattern2022limits}, the perturbed embeddings are used to generate anonymized texts that meet the requirements of local differential privacy. However, the added noise introduces grammatical errors and semantic deviations in the sanitized text, reducing its usability; \textbf{(2) Text Anonymization}\cite{lison2021anonymisation}: Some existing works treat privacy protection as a part-of-speech (POS) tagging task, utilizing named entity recognition (NER)\cite{nadeau2007survey} to capture, remove, or replace private information in the text. However, when entity information serves as a prerequisite for task execution, indiscriminate processing of named entity terms can render the prompts unusable. Additionally, users cannot directly use responses from cloud LLMs, as further de-anonymization is required\cite{chen2023hide}. Other works consider privacy protection as a text rewriting task, leveraging language models to rewrite the original text\cite{utpala2023locally}, but this only changes the writing style and does not obscure private information.

Due to high computational demands, extensive system modifications, limited application scenarios, low task usability, and coarse privacy protection, the aforementioned methods struggle to effectively address the issue of privacy leaks in the prompt. Therefore, we propose that a text-to-text prompt privacy protection framework for LLM applications should achieve the following three goals:

\textbf{(1) High Usability}: The framework should be applicable to various tasks, and the processed prompts should maintain high usability without affecting task accuracy. Since service providers may reject prompts with high perplexity to prevent jailbreak attacks\cite{zou2023universal}, anonymized prompts should also remain as fluent as possible.

\textbf{(2) Dynamic Anonymity}: The processed prompts should not contain private information while ensuring usability, and the protected privacy targets should change dynamically based on the context. For instance, the name `Jack' should only be considered a privacy leak in the sentence ``Jack is a billionaire." rather than in ``All work and no play makes Jack a dull boy."

\textbf{(3) Adaptivity}: This goal encompasses two aspects: \textit{(\romannumeral1) Audience Inclusivity}: The framework should be adjustable in terms of privacy protection performance and computational load requirements. This allows basic privacy protection for ordinary users with limited computing resources, while providing high-level anonymization of multiple data types for enterprises with abundant computing power. \textit{(\romannumeral2) System Independence}: The framework should be seamlessly integratable into mainstream NLP pipelines without requiring modifications to other components.

To meet the above goals, we propose the \texttt{\underline{Pro}mpt Privacy \underline{San}itizer} (i.e., \texttt{ProSan}), the first framework for prompt privacy protection that balances usability and anonymity, as shown in Fig. \ref{fig:1_1}. Unlike existing solutions, we leverage the LLM itself to quantify the impact of each word in the prompt on task usability, and selectively replace task-irrelevant word to ensure \textbf{usability}. Additionally, we use self-information to measure the sensitivity of words in context, dynamically adjusting \textbf{anonymity}. Furthermore, we utilize a masked language model to generate anonymized words based on context, enhancing fluency. Sampling probabilities of words are determined based on importance and self-information, achieving a dynamic balance between usability and anonymity. Moreover, we train a sequence-to-sequence (Seq2Seq) model on the generated text pairs, providing an out-of-the-box \texttt{ProSan} for ordinary users. Our framework operates at the text-to-text level, enabling seamless integration into existing LLM service pipelines, thus fulfilling \textbf{adaptivity} requirements.

We summarize the contributions as follows:
\begin{itemize}
\item We propose \texttt{ProSan}, an end-to-end framework for prompt privacy protection, which dynamically balances the usability and anonymity of prompts, ensuring the privacy of sensitive information while maintaining usability.
\item We combine the importance of words with self-information to flexibly identify and effectively replace sensitive words in prompts for the first time, providing a context-adaptable solution for different words rather than applying a uniform privacy protection strategy.
\item We treat privacy protection tasks as Seq2Seq tasks. By utilizing high-quality anonymized text pairs obtained from compute-intensive steps, we fine-tune a lightweight anonymized Seq2Seq model. This ensures that ordinary users can execute the privacy protection framework on local devices with minimal loss of task usability.
\item Empirically, our method presents an outstanding performance across multiple tasks. In question answering, text summarization, and code generation tasks, our approach accurately identifies and replaces sensitive information with almost no loss in usability.
\end{itemize}

\section{Background}
\subsection{Large Language Model and Prompt}
Language models learn statistical patterns of language based on training corpora, enabling them to predict and generate text that aligns with natural language conventions. The introduction of the Transformer architecture has significantly enhanced the ability of language models to handle long sequences and has improved computational efficiency\cite{vaswani2017attention}. Our framework involves three types of modern language models based on Transformer.

\textbf{Encoder-only models}, such as BERT\cite{devlin2019bert} and RoBERTa\cite{liu2019roberta}, focus on text comprehension. By understanding the semantic information of the input, they provide efficient text representations for downstream tasks like NER and sentiment classification.

\textbf{Decoder-only models}, such as GPT\cite{brown2020language} and LLaMA\cite{touvron2023llama}, emphasize text generation. They can generate sequences word by word based on given inputs, suitable for tasks like dialogue and question answering.

\textbf{Encoder-decoder models}, such as BART\cite{lewis2020bart} and Flan-T5\cite{chung2024scaling}, specialize in Seq2Seq tasks. The decoder generates corresponding output using context information captured by the encoder, applicable to translation and text summarization.

The substantial increase in the scale of model parameters and training data in LLMs has not only led to the phenomenon of emergent abilities\cite{wei2022emergent}, enabling models to understand instructions and engage in continuous contextual dialogues, but has also transformed NLP paradigms. From the initial fine-tuning\cite{devlin2019bert}, the field has evolved to few-shot learning\cite{brown2020language} and further to zero-shot prompting\cite{ouyang2022training}. Nowadays, with simple and intuitive prompts, people can guide LLMs to complete tasks, without the need to fine-tune model parameters or provide input-output demonstration pairs.

\subsection{Privacy Threats on LLMs}
With the rapid advancement of LLMs, their widespread application has raised concerns about the potential misuse of data. Throughout the lifecycle of LLMs, vast amounts of information are processed and generated, posing privacy leakage risks to various data subjects.

\textbf{Training Data}: There are many ways in which training data can be leaked. Due to the memory capabilities of LLMs\cite{tirumala2022memorization}, attackers can induce models to reveal sensitive personal information from training data by crafting specific query prefixes\cite{carlini2021extracting}. Attackers can also perform membership inference attacks to determine whether a text was part of the training data\cite{carlini2022membership}. In federated learning scenarios, even if participants only upload gradient updates, third parties can reconstruct the original training text based on these updates\cite{gupta2022recovering}. In both soft\cite{duan2024flocks} and discrete\cite{hong2023dp} prompt-tuning scenarios, private datasets utilized for tuning are vulnerable to the risk of unauthorized access.

\textbf{Inference Texts}: During the inference phase of LLMs, users inevitably expose their prompts to the model service provider to accomplish tasks. In the process, users' private information is collected by the provider and could be intercepted by third parties. Even if users use embeddings to prevent direct privacy leaks, attackers can still reconstruct original texts or train classifiers to infer specific private attributes from the embeddings\cite{pan2020privacy}. If users attempt to encrypt their prompts\cite{lin2024promptcrypt}, the model might not respond correctly, as most LLMs are not designed or trained to handle encrypted text.

We focus on privacy leakage during the \textbf{inference phase}. Even if data is transmitted in different forms, such as ciphertext or embedding, its private information does not disappear with changes in the medium. Therefore, we anonymize private information from the beginning when users disclose their data. Since the training corpora of LLMs primarily consist of user self-disclosures gathered from the Internet\cite{dou2023reducing}, our work also has positive implications for protecting the privacy of training data.

\subsection{Text Privacy Preservation}\label{subsec:Privacy-preservingOfText}
Prior works have focused on mitigating the risk of text privacy leaks by obscuring personal information. These efforts fall into two main categories.

\textbf{Paraphrase}: This strategy involves rephrasing the original text to reduce the risk of attackers tracing authorship through writing style or language patterns. Human-engineered methods use predefined rules, such as synonym substitution\cite{bevendorff2019heuristic} or word removal\cite{mahmood2019girl}, for paraphrasing. Deep learning methods utilize discriminator networks to penalize the generation of text that reveals the authorship attribution\cite{shetty2018a4nt}. However, paraphrasing often fails to hide crucial entity information related to personal privacy.

\textbf{Anonymization}: The goal of anonymization is to prevent privacy disclosure by removing personally identifiable information (PII), such as name, address, and phone number from the text. One approach involves POS tagging and techniques like regular expressions\cite{rosado2023pii} or NER\cite{herwanto2021named} to identify and remove PII. However, this method may fail when such information is an essential component of the prompt. Another approach employs privacy-preserving data publishing (PPDP) techniques like $k$-anonymity\cite{cumby2011machine} or differential privacy\cite{feyisetan2019leveraging}. Nevertheless, $k$-anonymity is limited to static datasets and does not apply to unstructured textual data. The use of differential privacy in text anonymization depends heavily on the choice of embedding models, which may prioritize statistical properties of the text over semantic features. Moreover, the introduced noise can reduce text usability. Such methods often provide a uniform mechanism for different words, tending to offer a formal privacy guarantee while overlooking real-world application scenarios.

\section{Problem Formulation}
\subsection{Threat Model}
We adhere to a typical scenario of using an LLM. Users prepare prompts locally and then send them to LLM service providers such as OpenAI and Google. The LLM completes the task online and sends the response back to the user. A typical prompt $q=\{i, x\}$ consists of an instruction $i$ describing the task the user wants the LLM to perform, such as ``Please diagnose based on the following symptoms:", and information $x$, which is the specific information, such as ``The patient's name is Liam. He has had a high fever of 39.5° for 2 days." The prompt may also contain user personal private information $\rho$ unrelated to the task, such as the name `Liam'. We assume that the LLM service providers are semi-honest, meaning they comply with agreements but have sufficient motivation to directly obtain or infer user private information $\rho$ from the prompt $q$, either for training data or improving service quality. Users, on the other hand, aim to remove the private information $\rho$ while ensuring the usability of the prompt.
\subsection{Design Goal}
Our goal is to construct a privacy-preserving framework for prompts in LLMs. However, applying previous relevant work directly to the current scenario presents the following challenges:

Solutions based on tagging may remove words from the information $x$ relevant to task execution, while solutions based on differential privacy introduce noise that decreases the readability of prompt $q$. Both approaches ultimately result in a reduction of usability. Although fine-tuning the language model on anonymized text can enhance performance\cite{qu2021natural}, in our scenario, LLM service providers would not modify their training pipelines for the sake of generality.

In previous works, the definition of privacy is based on predetermined rules or types, or relies on the choice of PPDP techniques. The measurement of privacy leakage risks depends on predetermined privacy budget or text embedding calculation methods. However, the definition and measurement of privacy need to be related to contextual integrity\cite{nissenbaum2004privacy}. The formal privacy guarantees provided by the above works do not adjust for different texts.

Previous works leverage online LLMs to protect the privacy information of text, but relying on untrusted service providers introduces a stronger assumption. Although it's possible to replace online LLMs with locally run open-source LLMs, the high computational cost brought by billions of parameters raises the barrier to users.

Considering the above issues, we aim for the \texttt{ProSan} to achieve the following objectives:
\begin{itemize}
\item \textbf{High Usability}: The framework treats lexical items as the basic processing units. The converted $\tilde{q}$ should retain readability and semantic content relevant to the task, enabling LLMs to respond correctly.
\item \textbf{Dynamic Anonymity}: The framework should dynamically assess the privacy leakage risk of each word based on the specific context and accordingly protect sensitive information.
\item \textbf{Adaptivity}: The framework can operate independently of online or computationally intensive LLMs and can run locally on the user side, such as mobile devices.
\end{itemize}

\section{Preliminary Analysis}\label{sec:PreliminaryAnalysis}
Our goal is to achieve dynamic privacy protection while maintaining usability. Therefore, before introducing the design of the \texttt{ProSan}, this section will explain how to quantify the impact of words in the prompt on task usability and the magnitude of privacy leakage risk.
\subsection{Measuring Utility Impact}\label{sec:MeasuringUtilityImpact}
Given a prompt $q$ consisting of $k$ words, $q = [w_1, w_2, \ldots, w_k]$, the tokenizer of an LLM decomposes each word into tokens, transforming it into $q = [t_1^{(1)}, t_2^{(1)}, \ldots, t_n^{(k)}]$, where $t_i^{(j)}$ denotes the $i$-th token belonging to the $j$-th word. The generation process of the LLM can be represented by the conditional probability $P_{\mathrm{LM}}^{T}(y \mid q)$, where $y$ is the corresponding output, and the temperature $T$ is used to adjust the diversity of $y$.

Model interpretability can help us understand how the word $w_j$ in the prompt affects the output $y$. In our framework, without access to the internal parameters of the cloud LLM, the most intuitive method is perturbation-based. That is, perturbing the $j$-th word $w_{j}$ in the prompt $q$, and calculating the absolute change in the loss function $\mathcal{L}$ of the perturbed prompt $q^{\prime}$ relative to the original $q$, as the importance score $K_{w_{j}}$ for $w_{j}$:

\begin{equation}
K_{w_{j}}=\left|\mathcal{L}(q, y)-\mathcal{L}(q^{\prime}, y)\right|
\end{equation}

But during implementation, we encountered the following issues:

(1) The perturbation operations are difficult to implement. The simplest way to perturb is to directly delete the word $w_{j}$, but this deletion may cause $q^{\prime}$ to deviate from the natural distribution of language, rendering it an invalid sentence. For instance, the importance of `San Francisco' lies in its entirety; removing `Francisco' would result in an invalid word, potentially inducing unexpected behavior in LLMs. Another approach is substitution, wherein $w_{j}$ is replaced with a word of similar type. However, finding a suitable replacement remains a formidable challenge.

(2) Capturing the interrelations between words is challenging. There are complex dependencies among the words in the prompt, and perturbing individual words may not accurately reflect their true importance. For example, in sentiment classification tasks, a sentence containing both `happy' and `joyful', independently deleting each word may not change the output. Obvious alteration in the output may only occur when both words are deleted simultaneously.

(3) High time complexity. A prompt of length $k$ requires at least $k$ perturbations to obtain importance scores for each word. If considering the combination features between words, the time complexity will exponentially increase with the sequence length. Under the constraints of response speed and frequency limits of existing LLM services, this problem is difficult to overcome.

Considering the various issues with perturbation-based methods, we opted for a gradient-based approach. It is relatively straightforward, has low time complexity, and resolves the non-differentiability issue by converting discrete text into token embeddings. Initially, we compute the gradient of the loss function $\mathcal{L}$ with respect to each token:
\begin{equation}
g_{t_{i}^{(j)}}=\frac{\partial \mathcal{L}(q, y)}{\partial t_{i}^{(j)}}
\end{equation}

As our framework operates at the lexical level, and in order to mitigate the influence of token count on the importance score, we calculate the gradient of each word using the following formula:
\begin{equation}
g_{w_{j}}=\frac{\sum_{i=a}^{a+\beta}g_{t_{i}^{(j)}}}{\beta}
\end{equation}
where $\left[t_{a}^{(j)}, t_{a+1}^{(j)}, \ldots, t_{a+\beta}^{(j)}\right]$ represents tokens corresponding to the word $w_j$. Finally, we compute the $l_2$ norm of $g_{w_{j}}$, followed by min-max normalization to derive a importance score $K_{w_{j}}$ for each word in the prompt:
\begin{equation}
K_{w_{j}}=\frac{\left\|g_{w_{j}}\right\|_{2}-\min_{i=1}^{k} g_{w_{i}}}{\max_{i=1}^{k} g_{w_{i}}-\min_{i=1}^{k} g_{w_{i}}}
\end{equation}

\begin{figure}[t]
\centering
\includegraphics[width=0.42\textwidth]{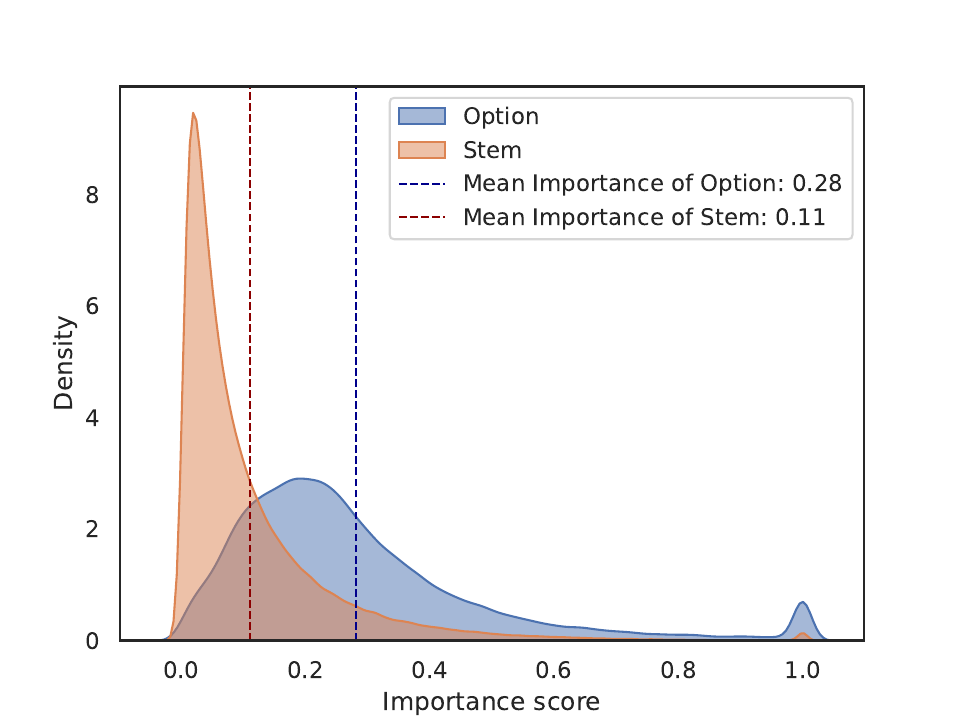}
\caption{The kernel density estimate plot of the importance scores for stems and options in multiple-choice questions from the MedQA dataset.}
\label{fig:4.1_1}
\end{figure}

Based on this method, we evaluated the word importance of multiple-choice questions in the MedQA dataset\cite{jin2020disease}. Each question consists of a stem and options. The kernel density estimate plot in Fig. \ref{fig:4.1_1} demonstrates a significant difference in the distribution of word importance between the stems and options. Options, serving as answer sources, generally have higher importance, indicating the effectiveness of this method.

Figure \ref{fig:4.1_2} visualizes the influence of each word in a prompt on the output `A'. It can be observed that words such as `chest pain', `dyspnea', and `coronary artery disease' play a crucial role in the decision-making process of the LLM by providing key evidence. Their importance scores are significantly higher than those of personal information such as `David', `Steve', and `New York City' which have minimal impact on the model's output despite also being proper nouns. This provides a foundation for identifying task-relevant words in subsequent steps to ensure the usability of prompts.

\begin{figure}[t]
    \centering
    \resizebox{\linewidth}{!}{
        \fbox{\parbox[c]{\linewidth}{ 
            {\setlength{\fboxsep}{-1pt}
                \input{texs/4.1_2.tex}
            }
        }}
    }
    \caption{A visualization of word importance within a prompt. Darker colors indicate greater importance.}
    \label{fig:4.1_2}
\end{figure}

Gradient-based methods necessitate access to the parameters of LLMs. While viable for users with ample computational resources or LLM trainers focused on training data anonymization, the high computational costs pose a significant barrier for ordinary users, even when considering open-source LLMs as alternatives. This is contradictory to the design goal of adaptivity. Our solution will be presented in \S \ref{sec:LightweightImplementation} .

\subsection{Measuring Privacy Risk}\label{sec:MeasuringPrivacyRisk}
Privacy information can be defined as information related to identified or identifiable individuals, including PII such as name, address, and phone number; it also includes quasi-identifiers, which cannot identify individuals in isolation but may reveal personal identity when combined with other information, such as gender, nationality, and race; it further encompasses personal sensitive information that is not suitable for public disclosure, such as occupation, sexual orientation, and health information.

One method for intuitively measuring privacy leakage risk is to predefine categories of protected words, but its static nature cannot adapt to the dynamic changes in privacy. Within the same category, the risk varies depending on the specific word. For instance, the privacy risk is higher for ordinary names compared to those of public figures like `Biden'. Moreover, the risk of the same word can fluctuate due to ambiguity and context. For example, in ``An apple a day keeps the doctor away." versus ``He is an employee of Apple Inc.," the latter's `Apple' reveals the data subject's employment information. 

Another approach is from the perspective of data publication, using word frequency, $t$-closeness, and differential privacy. These rely on factors such as corpus distribution, text embeddings, and privacy budgets to define privacy leak risks. However, these methods primarily aim to prevent attackers from inferring privacy via statistical data features, neglecting the fact that the core of text privacy lies in its semantics.

We tackle the measurement issue of privacy leakage risk from a fresh perspective, viewing the essence of privacy as information. By reducing the amount of information, we naturally mitigate the risk of leaks. This is also consistent with the principle of data minimisation in the GDPR\cite{europaRegulation2016679}. Therefore, we attempt to introduce the concept of self-information from information theory to quantify the risk of privacy leakage.

Self-information is used to measure the amount of information conveyed when an event occurs\cite{shannon2001mathematical}. It is obtained by calculating the negative log likelihood of the event's probability, expressed as $I(x)=-\log_2 p(x)$. The lower the probability of an event occurring, the higher its self-information. Self-information has been widely applied in accelerating\cite{li2023compressing} and evaluating\cite{bunescu2022distribution} LLMs. Its alternative term, `surprisal', aptly captures its essence: encountering an unlikely event brings about unexpected surprise. In the context of privacy protection, a text abundant in privacy content typically evokes higher surprisal, much like gossip filled with dramatic plots usually triggers greater surprise.

In the context of natural language modeling, events can be considered as the generation process of each token, and the probability precisely corresponds to its output distribution. Therefore, we use the following formula to calculate the self-information of each token $t_{i}$:
\begin{equation}
I(t_{i})=-\log_{2} P\left(t_{i} \mid t_{1}, t_{2}, \ldots, t_{i-1}\right)
\end{equation}

Due to the additivity of self-information, the self-information of the word $w_j$ can be defined as:
\begin{equation}
I_{w_{j}}=\sum_{i=a}^{a+\beta}I_{t_{i}^{(j)}}
\end{equation}

We also use min-max normalization to derive privacy scores $O_{w_{j}}$ for each word in the prompt:
\begin{equation}
O_{w_{j}}=\frac{I_{w_{j}}-\min_{i=1}^{k} I_{w_{i}}}{\max_{i=1}^{k} I_{w_{i}}-\min_{i=1}^{k} I_{w_{i}}}
\end{equation}

\begin{figure}[!t]
\centering
\includegraphics[width=0.42\textwidth]{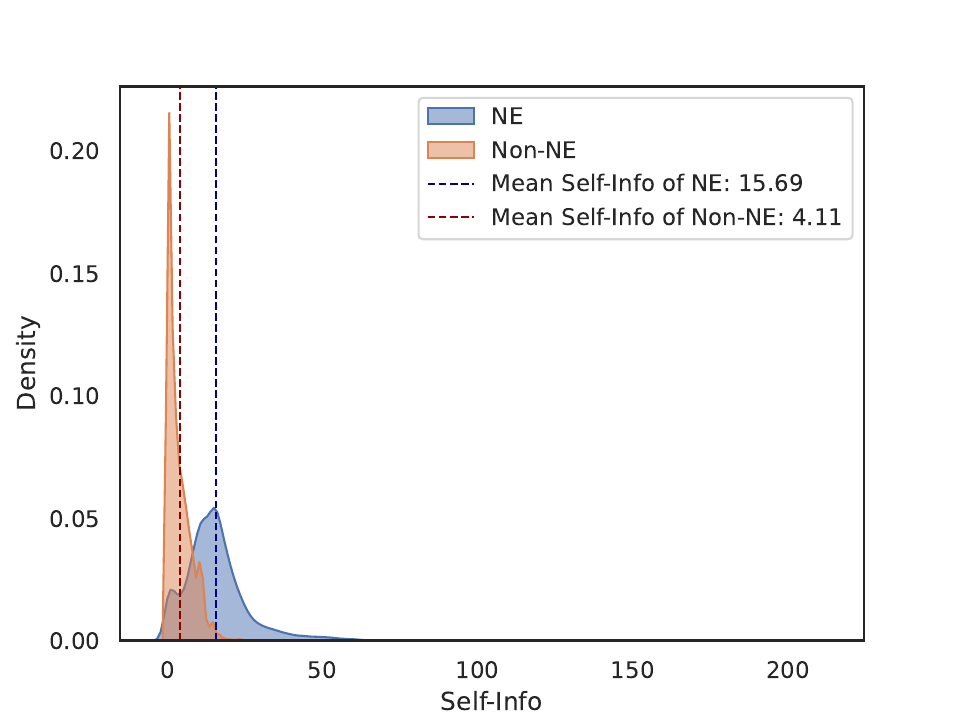}
\caption{The kernel density estimate plot of the self-information for named entities and non-named entities from the PII-masking-43k dataset.}
\label{fig:4.2_1}
\end{figure}

We utilize an autoregressive LLM to compute the conditional probability of generating each token, thus obtaining privacy scores for each word. When a token has a low conditional probability, and because the LLM is trained on a large-scale corpus, it indicates an infrequent occurrence of the token within the specified context in the corpus. Such tokens often contain more information, as their rarity in the desensitized corpus increases the likelihood of them carrying privacy information. To further validate our hypothesis, we randomly sampled 10,000 sentences from the PII-masking-43k dataset\cite{ai4privacy_2023}, categorizing words into named entities (NE) and non-named entities (Non-NE), and computed the self-information of each word. Figure \ref{fig:4.2_1} illustrates the kernel density estimation plot of self-information for NE and Non-NE, revealing a significant difference in their distributions. Named entities, typically associated with higher privacy leakage risks, exhibit higher self-information.

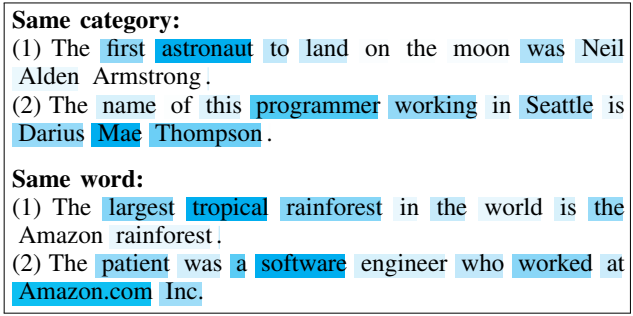
\begin{figure}[!t]
    \centering
    \resizebox{\linewidth}{!}{
        \fbox{\parbox[c]{\linewidth}{ 
            {\setlength{\fboxsep}{-1pt}
                \textbf{Same category: }\\
                (1) \input{texs/4.2_2.tex}\\
                (2) \input{texs/4.2_3.tex}
                \vskip 0.1in
                \textbf{Same word: }\\
                (1) \input{texs/4.2_4.tex}\\
                (2) \input{texs/4.2_5.tex}
            }
        }}
    }
    \caption{A visualization of privacy leak risks for same category and same word. Darker colors indicate higher privacy leakage risk.}
    \label{fig:4.2_2}
\end{figure}

Figure \ref{fig:4.2_2} illustrates the dynamic aspect of the privacy measurement method based on self-information. In two sentences where the positions of the names `Neil Alden Armstrong' and `Darius Mae Thompson' are the same, the former, being a public figure, carries lower privacy leakage risk. Due to the ambiguity of `Amazon', only the occurrence in the second sentence leaks employment information, thus resulting in a higher privacy leak risk. These examples demonstrate our method's adaptability in assessing privacy risks based on different contexts. Our solution for concealing semantic features in the text will be presented in \S \ref{sec:TheProcessingofWords}.

\section{Design}
We will rely on the importance and privacy of words in prompts, using masked language models and Seq2Seq language models to implement end-to-end privacy protection for prompts. Our framework takes words as the basic unit of privacy protection, and this section will detail our design around the following three core questions:
\begin{itemize}
\item Which words in a prompt should be selected for privacy protection?
\item What methods should be used to process these words?
\item How can we make the anonymous process feasible for users lacking computational resources?
\end{itemize}

\subsection{The Choice of Words}
Considering efficiency and the fact that not all words pose privacy risks, it's unnecessary to protect every word. We'll select words for privacy protection based on three dimensions: the privacy risk associated with the prompt, the importance of the word, and the word's POS.

The privacy leakage risks vary for different prompts, and stricter privacy protection is needed for prompts with higher risks. We calculate the average self-information $\frac{1}{n}\sum_{j=1}^{n}I_{w_j}$ of the word in the prompt $q$, which is also the entropy $H_q$ of the prompt, as a metric for measuring the privacy leakage risk of the prompt. As discussed in \S \ref{sec:MeasuringPrivacyRisk}, higher entropy implies more privacy risk. We adaptively adjust the proportion $\gamma_{p}$ of words in the current prompt that need to be protected using the following formula:
\begin{equation}
\gamma_{q} = \lambda \frac{1}{1+e^{-H_{p}}}
\end{equation}
where $\lambda$ is the scaling factor.

Once the protection ratio is determined, we will sort each word according to the importance scores in ascending order. The set of words $\mathcal{W}^{*}$ within the first $\gamma_{q}$ will be our primary focus. Due to the relatively lower importance scores of these words, processing them minimally impacts the model's response accuracy, aligning with our usability design goal. We don't prioritize handling words with the highest privacy scores directly, as prompts may rely on proper nouns, and even minor disruptions to them can significantly impact usability.

Even within $\mathcal{W}^{*}$, not all words require processing. We utilize \texttt{spaCy}\footnote{\url{https://spacy.io}} to identify word types. Function words such as prepositions, conjunctions, and articles primarily serve grammatical roles rather than conveying specific semantics. Therefore, our focus lies more on content words such as nouns (including proper nouns), numbers, and adjectives. These words typically carry explicit semantic meanings and directly convey information. Unlike privacy protection methods focusing solely on NE like name and email, our approach can obscure Non-NE such as occupation, skin color, and sexual orientation, preventing tracing attacks by adversaries using quasi-identifiers or writing styles. Our framework allows customization of word types of interest, yielding the final set of words to be processed, denoted as $\mathcal{W}^{\prime}$ after POS filtering.

\begin{figure*}[t]
\centering
\includegraphics[trim=22mm 9mm 27mm 12mm, clip, width=0.87\textwidth]{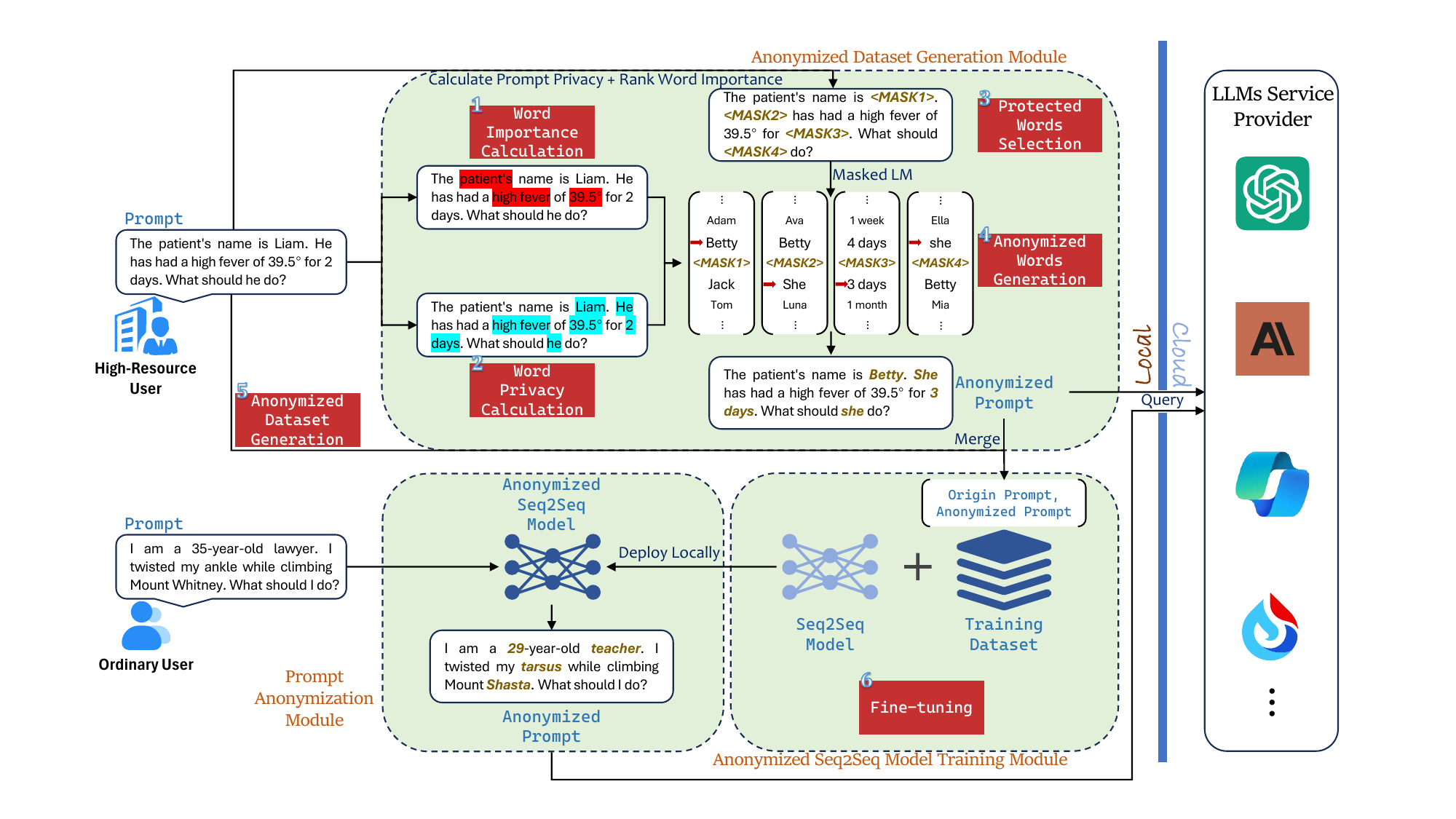}
\caption{\texttt{ProSan} overview. (1) For high-resource users, they use LLMs to assess the importance and privacy of each word in the prompt. Based on overall privacy risk and word importance, they select words to protect, use a masked language model to predict candidate words, and determine sampling probabilities by combining word importance and privacy. This process results in customizable anonymized prompts. (2) For trusted third parties, they use high-quality \textit{\textless original prompt, anonymized prompt \textgreater} pairs generated by the Anonymized Dataset Generation Module to fine-tune existing Seq2Seq models, obtaining anonymized models which are then released publicly. (3) For ordinary users, they deploy the anonymized model locally to achieve basic, fast, end-to-end prompt anonymization by simply inputting their prompts.}
\label{fig:5.4_1}
\end{figure*}

\subsection{The Processing of Words}\label{sec:TheProcessingofWords}
Deleting words from $\mathcal{W}^{\prime}$ is the most direct way to remove privacy. However, it may cause grammatical errors in prompts. Therefore, we consider using a substitution strategy to handle words. Previous approaches relied on word embedding, NE types, or the hash value of the original text to generate replacement words. However, due to the lack of context, these replacements may distort prompt semantics. We utilize a masked language model to generate replacement words. These models are trained by predicting the randomly masked words in the input text, allowing them to incorporate contextual information when generating replacement words, ensuring semantic appropriateness and coherence.

The direct use of words predicted by the masked language model as replacements poses two potential risks. One is when the replacement word is too similar in meaning to the original, rendering privacy protection ineffective. The other is when the replacement word deviates too much from the original word's meaning, potentially affecting the accuracy of the LLM output. To dynamically balance privacy and usability, we jointly determine replacement word generation based on three dimensions: word importance score, privacy score, and similarity to the original word.

For each word in $\mathcal{W}^{\prime}$ requiring protection, we initially mask it with \texttt{<MASK>} in the original text. Then, we use a masked language model to predict multiple candidate words $c_i$ and their corresponding probabilities $p_i$ at that position. We sort the candidate words in descending order of probability and accumulate probabilities until reaching $\tau$. All words before reaching this threshold constitute the candidate set $\mathcal{C}_{w^{\prime}}$ for the word $w^{\prime}$, defined as follows:
\begin{equation}
\begin{gathered}
\mathcal{C}_{w^{\prime}} = \{c_1, c_2, \ldots, c_k \mid \sum_{i=1}^{k-1}p_i < \tau \leq \sum_{i=1}^{k}p_i,
\\
p_1 \geq p_2 \geq \ldots \geq p_n,1 \leq i \leq n \}
\end{gathered}
\end{equation}

After obtaining the candidate word set $\mathcal{C}_{w^{\prime}}$, we then recalculate the sampling probability $p_i^{\prime}$ for each candidate word using the following formula:
\begin{equation}\label{equ:p}
p_{i}^{\prime}=\frac{e^{(K_{w^{\prime}}-O_{w^{\prime}}) \cdot s_{i}}}{\sum_{j=1}^{k} e^{(K_{w^{\prime}}-O_{w^{\prime}}) \cdot s_{j}}}
\end{equation}

Here, $K_{w^{\prime}}$ and $O_{w^{\prime}}$ respectively represent the importance score and privacy score of the protected word $w^{\prime}$, while $s_i$ denotes the similarity between candidate word $c_{i}$ and $w^{\prime}$, with higher values indicating greater similarity. Equation \ref{equ:p} reveals that for words with significant output impact ($K_{w^{\prime}}$ high) and low privacy risk ($O_{w^{\prime}}$ small), such as task-relevant generic nouns or specialized terms, candidate words closer to $w^{\prime}$ are favored to maintain text usability. Conversely, for words with minor output impact ($K_{w^{\prime}}$ low) and higher privacy risk ($O_{w^{\prime}}$ large), such as irrelevant personal information or company confidential data, candidate words differing more from $w^{\prime}$ are prioritized to enhance privacy protection.

For computing importance and privacy scores, we have elaborated in \S \ref{sec:PreliminaryAnalysis}. Regarding measuring word similarity, existing methods mostly utilize word embedding techniques like word2vec\cite{mikolov2013efficient}, GloVe\cite{pennington2014glove}, or ELMo\cite{peters2018deep}. These techniques train on large-scale corpora to map semantic information of words into multi-dimensional vectors, assessing word similarity through cosine similarity between vectors. While these methods based on skip-gram models, CBOW models, or co-occurrence matrices effectively capture common usage patterns of words in specific contexts, they may not fully reflect semantic relations between words. For instance, in the GloVe model, the cosine similarity between `America' and its synonym `USA' is only 0.65, lower than its 0.72 similarity with `Canada'.

In order to accurately capture the semantic relationships between words in semantically-driven prompting-based privacy protection, we utilize \texttt{WordNet}\cite{miller1995wordnet}, a large lexical database that organizes words into synsets, sets of synonyms. In WordNet, different synsets are connected by semantic relations such as synonymy/antonymy, hypernymy/hyponymy. For a candidate word $c_{i}$ and the original word $w^{\prime}$, we calculate the similarity $s_i$ by computing the shortest path length between their respective synsets, ranging from 0 to 1. This approach yields a reasonable similarity score of 1 between `America' and `USA', whereas the similarity with `Canada' is only 0.33.

We assign new probabilities to each candidate word using Eq. \ref{equ:p} and select a word from $\mathcal{C}_{w^{\prime}}$ based on these probabilities to replace $w^{\prime}$. By repeating this process for all words in $\mathcal{W}^{\prime}$, we obtain the anonymized prompt $\tilde{q}$.
\subsection{Lightweight Implementation}\label{sec:LightweightImplementation}
While we have obtained the anonymized prompt $\tilde{q}$ that satisfies both usability and dynamic anonymity, this process relies on LLMs to compute the importance and privacy scores of words. This does not require enterprise users to have the capability to run commercial LLMs like ChatGPT; instead, it only requires a few consumer-grade GPU cards, enabling them to run existing open-source billion-parameter LLMs and thereby meet the operational requirements. However, for ordinary users with more urgent needs for prompt privacy protection, bearing such high computational costs is still impractical.

To tackle this issue, we treat the prompt privacy protection task as a Seq2Seq task. In the field of NLP, many tasks boil down to Seq2Seq, such as translation (\textit{\textless source language \textgreater} $\rightarrow$ \textit{\textless target language \textgreater}), text summarization (\textit{\textless original text \textgreater} $\rightarrow$ \textit{\textless summary \textgreater}), and question answering (\textit{\textless question \textgreater} $\rightarrow$ \textit{\textless answer \textgreater}). In such a framework, the privacy protection task can be represented as a sequence transformation process from \textit{\textless original prompt \textgreater} to \textit{\textless anonymized prompt \textgreater}.

Compared to LLMs with hundreds of billions of parameters, there are already many Seq2Seq language models with millions of parameters, and existing technologies enable models of this scale to run smoothly on mobile devices\cite{ge2022edgeformer}. Such models achieve outstanding performance on various Seq2Seq tasks, largely due to fine-tuning on large-scale, high-quality training datasets. For mainstream NLP tasks such as translation, text summarization, and reading comprehension, there already exist many high-quality datasets. However, existing privacy protection tasks are often treated as annotation tasks for privacy keywords, providing mostly the position of privacy terms in the original text, while ignoring the availability of the text for downstream tasks. Furthermore, there is a lack of objective metrics for evaluating the effectiveness of text privacy protection. Therefore, in the field of privacy protection, datasets for transforming original prompts into anonymized prompts are scarce.

Fortunately, we have already presented an automated prompt anonymization scheme in the previous section, enabling us to generate massive \textit{\textless original prompt, anonymized prompt \textgreater} pairs as training data. We have trained the existing Seq2Seq model on this dataset and released the fine-tuned model publicly. Users only need to input the original prompt $q$ into this model to obtain a usable anonymized prompt $\tilde{q}$. The entire process does not require computing the importance and privacy scores of words using LLMs or generating replacement words using masked language models, greatly reducing the computational burden on the user side and thus meeting the design goal of adaptivity.

\subsection{Framework Summary}
After addressing the three questions mentioned above, we will present the summary of \texttt{ProSan}. As shown in Fig. \ref{fig:5.4_1}, \texttt{ProSan} consists of three modules.

\textbf{(1) Anonymized Dataset Generation Module}: As this module requires running a billion-parameter-level open-source LLM locally, it primarily provides customizable prompt anonymization services for users with abundant computational resources, such as enterprises, governments, and educational institutions. It is also used for generating anonymized datasets. Initially, the module uses the LLM to evaluate each word's impact on the output (importance score) and its privacy risk (privacy score). Then, it decides the protection proportion based on the overall privacy risk level of the prompt, and selects the words needing protection according to their importance scores. Next, a masked language model generates candidate replacements for the words to be protected, ensuring the fluency of the prompt. Finally, it selects the replacement words based on both the importance and privacy of the original words, replacing them to produce the final anonymized prompt.

\textbf{(2) Anonymized Seq2Seq Model Training Module}: This module fine-tunes the Seq2Seq model using text pairs consisting of original and anonymized prompts (generated by the Anonymized Dataset Generation Module). Once training is complete, the model will be released to the public.

\textbf{(3) Prompt Anonymization Module}: As this module only contains a Seq2Seq model with millions of parameters, it primarily offers basic prompt anonymization services for ordinary users with limited computational resources. Users can simply input the prompt to be protected into the anonymized Seq2Seq model locally, and immediately obtain an anonymized prompt.

\section{Evaluation}
In this section, we evaluated the performance of the \texttt{ProSan} in terms of usability and anonymity through a series of experiments. Our evaluation also analyzed its performance variation under different computational resources to verify its adaptivity. We choose two existing text anonymization solutions, HaS\cite{chen2023hide} and SanText+\cite{yue2021differential}, as baselines and conducted comprehensive comparisons with our framework in three representative tasks. Furthermore, we performed ablation experiments to explore the impact of each module in the \texttt{ProSan} on performance.

\subsection{Experienment Setup}
\subsubsection{Datasets}
We chose three different datasets to assess the performance of \texttt{ProSan}, representing three typical usage scenarios of online LLMs: question answering, text summarization, and code generation. For question answering, we used the MedQA\cite{jin2021disease}, comprising real-world medical questions and corresponding answers; for text summarization, we employed the SAMSum\cite{gliwa2019samsum}, consisting of messenger-like conversations with corresponding summaries; and for code generation tasks, we utilized the CodeAlpaca\cite{iamtarun_2023}, which generates instructional tasks, input examples, and solution codes through Self-Instruct\cite{wang2023self}.

However, these datasets were desensitized before release, lacking annotations for privacy information. Moreover, existing privacy evaluation datasets are not tailored for the prompt scenario. Thus, to evaluate the usability and privacy of anonymized prompts, we artificially inserted privacy information into the original prompts from the following perspectives:

\textbf{(1) Non-interference}: The inserted privacy does not affect the original prompt's purpose and output. We added patient privacy information unrelated to answers in the MedQA, transcriber privacy information unrelated to summaries in the SAMSum, and confidential company information unrelated to codes in the CodeAlpaca, to assess the protection capabilities for personal privacy and company confidentiality of \texttt{ProSan}. Table \ref{tab:changes} indicates that the inserted privacy has minimal impact on the original prompts' usability.

\begin{table}[ht]
\caption{Changes in the usability of the dataset before and after insertion, as reflected in the performance of the corresponding fine-tuned LLaMA-2-7B.}
\label{tab:changes}
\centering
\begin{tabular}{ccccc}
\toprule[1.5pt]
\bf Dataset & \bf Metric & \bf Original & \bf Inserted & \bf $\Delta_{\text{Utility}}$\\ 
\toprule
MedQA & Accuracy & 0.571 & 0.573 & +0.002 \\
\midrule
SAMSum & RougeL & 0.303 & 0.299 & -0.004\\
\midrule
CodeAlpaca & CodeBLEU & 0.283 & 0.280 & -0.003 \\
\bottomrule[1.5pt]
\end{tabular}
\end{table}

\textbf{(2) Hard-to-capture}: The inserted privacy is not easily captured. Therefore, we chose to insert privacy attributes including Non-NE attributes such as race, occupation, and sexual orientation, as well as highly customized username and password. Additionally, some inserted privacy attributes already exist in the original prompt. For instance, after insertion, the SAMSum would contain interlocutor names highly correlated with the answers and unrelated transcriber names. Therefore, rule-based solutions such as regular expressions or NER struggle to capture the privacy we added, and even if they do, it would decrease the usability of the prompts. Moreover, the syntax and insertion positions of privacy statements are randomly generated. We have provided the details of the datasets processed in Tab. \ref{tab:details}.

\begin{table}[ht]
\caption{Details of the datasets.}
\label{tab:details}
\centering
\begin{tabular}{
  >{\centering\arraybackslash}m{0.15\columnwidth}
  >{\centering\arraybackslash}m{0.2\columnwidth}
  >{\centering\arraybackslash}m{0.1\columnwidth}
  >{\centering\arraybackslash}m{0.35\columnwidth}}
\toprule[1.5pt]
\bf Dataset & \bf Task & \bf Size & \bf Inserted Privacy Attributes \\ 
\toprule
MedQA & Q\&A & 10,200 & Name, Occupation, Sexual orientation, Age \\
\midrule[0.5pt]
    SAMSum & Summarization & 16,369 & Name, Race, Location, Date \\
\midrule[0.5pt]
CodeAlpaca & Code generation & 18,000 & Username, Password, Corporate name, System configuration\\
\bottomrule[1.5pt]
\end{tabular}
\end{table}

\subsubsection{Baselines}
We have selected HaS\cite{chen2023hide} and SanText+\cite{yue2021differential} as comparison methods, which respectively represent the two approaches to text anonymization discussed in \S \ref{subsec:Privacy-preservingOfText}: based on POS tagging and PPDP.
\begin{itemize}
\item \textbf{HaS}: In the training phase, the framework utilizes an LLM to generate a dataset containing lexical POS tagging information and corresponding substituted words, which is used to train the Hide-model and Seek-model. In the inference phase, users anonymize a prompt using the locally run Hide-model and submit it to the cloud LLM. The response is then de-anonymized by the Seek-model to obtain the correct results.
\item \textbf{SanText+}: The framework calculates the sampling probabilities of substituted words based on word embedding distances and utilizes differential privacy to ensure the privacy of the sanitized words. Additionally, it provides tailored replacement strategies based on word usage frequency to enhance usability.
\end{itemize}

For HaS, we evaluate privacy using the anonymized prompt from the Hide-model, and assess usability with the results de-anonymized by the Seek-model. Consistent with \texttt{ProSan}, it also uses the `en\_core\_web\_sm' spaCy model. For SanText+, we use GloVe to obtain word embeddings and set the privacy parameter $\epsilon=15$, while all other configurations follow the implementation provided in the respective Github repositories\footnote{\url{https://github.com/alohachen/Hide-and-Seek}}\footnote{\url{https://github.com/xiangyue9607/SanText}}.
\subsubsection{Evaluation Metrics}
For privacy, we focus on the \textbf{Privacy Hiding Rate (PHR)}, calculated as follows:
\begin{equation}\label{equ:phr}
\text{PHR}=\frac{N_{hidden}}{N_{total}}
\end{equation}
where $N_{total}$ is the total number of privacy items in the prompts, and $N_{hidden}$ is the number of privacy items successfully removed in the anonymized prompts. For privacy attributes such as name, location, and company name, as well as those containing numbers like age, date, or password, replacing them is considered sufficient to remove privacy. However, for privacy attributes such as occupation, sexual orientation, and race, privacy risks persist even with non-identical word replacements if the semantics are similar. Therefore, we consider the semantics of such attributes. In \texttt{WordNet}, synonyms typically have a similarity above 0.9. Thus, only replacements with a similarity below 0.9 to the original word are counted in $N_{hidden}$. A higher PHR indicates stronger privacy protection.

For usability, we submit prompts to LLMs both before and after anonymization and assess usability based on the changes in the model's responses. The smaller the decrease, the higher the usability. Different tasks have different evaluation metrics. For MedQA, we use \textbf{Accuracy}, evaluating answer quality by calculating the proportion of correct responses. For SAMSum, we use \textbf{RougeL}, assessing summary quality by measuring the longest common subsequence between the generated summary and the reference summary. For CodeAlpaca, we use \textbf{CodeBLEU}, measuring code generation quality by comparing the syntactic and semantic similarities between the generated code and the reference code. Additionally, to prevent LLM providers from rejecting unreadable prompts and to enable users to quickly verify the effective removal of privacy information, we evaluate the fluency of anonymized prompts using \textbf{Perplexity}, lower perplexity indicates higher fluency.

\begin{table*}[ht]
\caption{The anonymity performance of anonymization methods(\%).}
\label{tab:anonymity}
\centering
\begin{tabular}{c|c|>{\centering\arraybackslash}m{0.21\columnwidth}|>{\centering\arraybackslash}m{0.21\columnwidth}|>{\centering\arraybackslash}m{0.21\columnwidth}|>{\centering\arraybackslash}m{0.21\columnwidth}}
\toprule[1.5pt]
    \bf Task & \bf \multirow{2}{*}{Privacy Attribute} & \multirow{2}{*}{HaS} & \multirow{2}{*}{SanText+} & \bf \multirow{2}{*}{$\text{ProSan}^{\diamond}$} & \bf \multirow{2}{*}{$\text{ProSan}^{\ast}$} \\
\bf (Dataset) & ~ & ~ & ~ & ~ & ~ \\
\toprule
\multirow{4}{*}{\begin{tabular}{c}Question Answering \\(MedQA) \\\end{tabular}} & Name & 99.9 & 51.5 & 97.0 & 99.2 \\
~ & Occupation & 1.3 & 65.9 & 92.2 & 100 \\
~ & Sexual Orientation & 1.5 & 61.5 & 92.7 & 99.7 \\
~ & Age & 99.8 & 100 & 98.3 & 100 \\
\midrule[0.5pt]
\multirow{4}{*}{\begin{tabular}{c}Text Summarization \\(SAMSum) \\\end{tabular}} & Name & 100 & 47.6 & 94.7 & 98.9 \\
~ & Race & 84.2 & 46.1 & 91.3 & 100 \\
~ & Location & 100 & 61.5 & 97.4 & 100 \\
~ & Date & 95.6 & 100 & 97.1 & 100 \\
\midrule[0.5pt]
\multirow{4}{*}{\begin{tabular}{c}Code Generation\\(CodeAlpaca) \\\end{tabular}} & Username & 9.2 & 29.5 & 100 & 100 \\
~ & Password & 7.8 & 99.7 & 100 & 100 \\
~ & Corporate Name & 97.7 & 43.9 & 92.8 & 100 \\
~ & System
Configuration & 3.1 & 95.2 & 97.9 & 99.7 \\
\midrule[0.5pt]
\midrule[0.5pt]
\multicolumn{2}{c|}{\bf Overall Privacy Hiding Rate} & 58.3 & 66.9 & 96.0 & \bf 99.8 \\
\bottomrule[1.5pt]
\end{tabular}
\end{table*}

\subsubsection{Implementation Details}
We use LLaMA-2-7b to calculate the importance and privacy scores of words in prompts, and generate replacement words using RoBERTa-base. The anonymized Seq2Seq model is trained on Flan-T5-base, with a learning rate of $\eta=1 \times 10^{-4}$ and weight decay of $\lambda=1 \times 10^{-2}$. For usability evaluation, we selected OpenAI's ChatGPT-3.5 and iFlytek's Spark-3.5 as the online LLMs for testing, interacting via APIs at a temperature setting of $T=1 \times 10^{-3}$. To mitigate fluctuations in experimental results due to cloud service iterations, we also conduct tests locally using an open-source LLaMA-2-7b that has been fine-tuned on the respective original datasets. We randomly extract 1000 samples from each dataset as test set, which is used to evaluate the overall performance of \texttt{ProSan} under both resource-rich (Anonymized Dataset Generation Module) and resource-restricted (Prompt Anonymization Module) conditions compared to baselines. The remaining data are fed into the Anonymized Dataset Generation Module to generate a dataset of \textit{\textless original prompt, anonymized prompt \textgreater} for fine-tuning the anonymized Seq2Seq model.

\begin{table}[ht]
\caption{The usability performance of anonymization methods.}
\label{tab:usability}
\centering
\begin{tabular}{@{}c|c|ccc@{}}
\toprule[1.5pt]
\bf Task & \bf \multirow{2}{*}{Method} & \multirow{2}{*}{GPT-3.5} & \multirow{2}{*}{Spark-3.5} & \multirow{2}{*}{LLaMA-2-7b} \\
\bf (Metric) & ~ & ~ & ~ & ~ \\
\toprule
\multirow{5}{*}{\begin{tabular}{@{}c@{}}Question Answering \\(Accuracy) \\\end{tabular}} & Baseline & 0.521 & 0.434 & 0.573 \\
\cmidrule{2-5}
~ & HaS & 0.201 & 0.230 & 0.215 \\
~ & SanText+ & 0.205 & 0.245 & 0.177 \\
~ & \bf $\text{ProSan}^{\diamond}$ & \bf 0.498 & \bf 0.430 & \bf 0.551 \\
~ & \bf $\text{ProSan}^{\ast}$ & 0.476 & 0.413 & 0.527 \\
\midrule[0.5pt]
\multirow{5}{*}{\begin{tabular}{@{}c@{}}Text Summarization \\(RougeL) \\\end{tabular}} & Baseline & 0.307 & 0.321 & 0.299 \\
\cmidrule{2-5}
~ & HaS & 0.213 & 0.216 & 0.207 \\
~ & SanText+ & 0.126 & 0.119 & 0.116 \\
~ & \bf $\text{ProSan}^{\diamond}$ & \bf 0.311 & \bf 0.318 & \bf 0.287 \\
~ & \bf $\text{ProSan}^{\ast}$ & 0.291 & 0.300 & 0.278 \\
\midrule[0.5pt]
\multirow{5}{*}{\begin{tabular}{@{}c@{}}Code Generation\\(CodeBLEU) \\\end{tabular}} & Baseline & 0.312 & 0.317 & 0.280 \\
\cmidrule{2-5}
~ & HaS & 0.293 & 0.289 & 0.242 \\
~ & SanText+ & 0.162 & 0.185 & 0.151 \\
~ & \bf $\text{ProSan}^{\diamond}$ & 0.295 & 0.307 & \bf 0.268 \\
~ & \bf $\text{ProSan}^{\ast}$ & \bf 0.312 & \bf 0.317 & 0.258 \\
\bottomrule[1.5pt]
\end{tabular}
\end{table}

\subsection{Overall Performance}
\subsubsection{Privacy}Table \ref{tab:anonymity} shows the result of the comparison between \texttt{ProSan} and the baseline. $\text{ProSan}^{\diamond}$ and $\text{ProSan}^{\ast}$ represent anonymized prompts generated by the Anonymized Dataset Generation Module and the Prompt Anonymization Module, respectively. We found that \texttt{ProSan} achieved high PHR across all privacy attributes. $\text{ProSan}^{\diamond}$, from the perspective of self-information, dynamically adjusts the strength and targets of privacy protection, achieving an overall PHR of 96.0\%. With high-quality anonymized text pairs, $\text{ProSan}^{\ast}$ achieved an even better overall PHR of 99.8\%.

However, SanText+, which relies on preset privacy parameters to adjust the strength of privacy protection, can not achieve acceptable performance with non-numeric privacy attributes. Similarly, HaS, which relies on predefined POS to determine privacy protection targets, does not perform well for non-entity privacy attributes. Although SanText+ and HaS can protect certain attributes effectively, they often result in significant usability loss. We will analyze these issues in detail in \S \ref{subsubsec:usability}.

\begin{figure}[!t]
    \centering
    \resizebox{\linewidth}{!}{
        \fbox{\parbox[c]{\linewidth}{ 
            {\setlength{\fboxsep}{-1pt}
                \textbf{Original prompt: }\\
                Q: A \bluecolorize{70}{30-year-old} \bluecolorize{70}{farmer} named \bluecolorize{70}{James} presents with an \redcolorize{70}{aortic diameter of 5.0 cm}. Genetic testing confirms \redcolorize{70}{Marfan Syndrome}. What is the best medication for his condition? \{`A': `Metformin', `B': `Lisinopril', `C': `\redcolorize{70}{Atenolol}', `D': `Albuterol', `E': `Prednisone'\}\\
                \textbf{HaS: }\\
                Q: A \reduline{45}-year-old farmer named \reduline{John} presents with an aortic diameter of \reduline{3.0} cm. Genetic testing confirms \reduline{Elmo's} Syndrome. What is the best medication for his condition? \{`A': `\reduline{Glucophage}', `B': `\reduline{Warfarin}', `C': `\reduline{Clopidogrel}', `D': `\reduline{Inotrofin}', `E': `\reduline{Sparanoid}'\}\\
                \textbf{SanText+: }\\
                \reduline{condition} : A \reduline{farmer What} year - \reduline{? for} named James presents \reduline{. confirms} aortic \reduline{cm} of \reduline{\}} cm \reduline{for} Genetic testing confirms \reduline{testing} Syndrome . What \reduline{James} the best medication \reduline{the} his \reduline{an of presents Prednisone} A ' : ` Metformin \reduline{?} , \reduline{of} B ' : ` Lisinopril ' , ` \reduline{year} ' : ` Atenolol ' , ` \reduline{of} ' \reduline{confirms} ` Albuterol ' \reduline{B} ` E ' \reduline{C} ` Prednisone ' \reduline{\{}\\
                \textbf{ProSan: }\\
                Q: A \reduline{70-ye}-old \reduline{patient} named \reduline{Robert} presents with an aortic diameter of 5.0 \reduline{centimeters}. Genetic \reduline{information} confirms Marfan Syndrome. What is the best medication for his \reduline{diagnosis}? \{`A': `Metformin', `B': `Lisinopril', `C': `Atenolol', `D': `Albuterol', `E': `\reduline{Others}'\}
            }
        }}
    }
    \caption{Comparison of anonymized prompts processed by different methods on the same prompt.}
    \label{fig:6.2.2_1}
\end{figure}

\subsubsection{Usability}\label{subsubsec:usability}
\textbf{Q\&A}. Table \ref{tab:usability} shows that the accuracy of $\text{ProSan}^{\diamond}$ decreased by only 0.4\%--2.3\%, and even the resource-limited $\text{ProSan}^{\ast}$ decreased by just 2.1\%--4.6\%, which is acceptable. In contrast, HaS and SanText+ drops by up to 18.9\%--39.6\%. HaS and SanText+, due to their overly broad or rigid protection goals, provide high protection for certain privacy attributes but inevitably disrupt important words. We analyze the anonymized results of different methods on the same prompt from Fig. \ref{fig:6.2.2_1}. In this figure, red highlights indicate crucial information for the model's decisions, blue highlights are irrelevant privacy information, and red underlines show the modifications by each method. We can see that HaS focuses on hiding entity information, which, while protecting privacy, also removes information about conditions, diseases, and medications, thus reducing usability. SanText+, lacking clear protection goals, leaks some privacy information while replacing certain keywords.

\begin{figure}[t]
\centering
\includegraphics[trim=0mm 6mm 0mm 8mm, clip, width=0.35\textwidth]{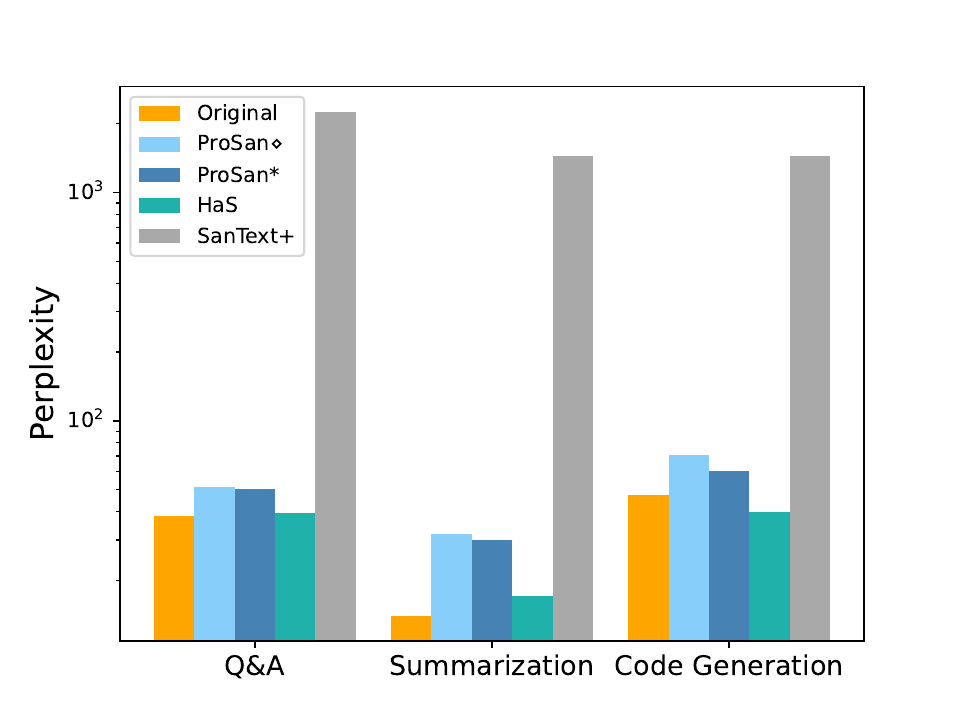}
\caption{The perplexity of prompts anonymized by different methods. Lower perplexity indicates higher fluency.}
\label{fig:6.2.2_2}
\end{figure}

\textbf{Summarization}. As shown in Tab. \ref{tab:usability}, the RougeL scores of $\text{ProSan}^{\diamond}$ and $\text{ProSan}^{\ast}$ decreased by no more than 0.012 and 0.021, respectively. In contrast, SanText+ experienced a decline of 0.181--0.202 due to the noise added during anonymization that caused semantic errors, as Fig. \ref{fig:6.2.2_2} illustrates with increased perplexity of anonymized prompt processed by SanText+. HaS saw a drop of 0.092--0.105. The SAMSum dataset includes both interlocutor names (essential for summaries) and transcriber names (private information), requiring an anonymization method that preserves the former for summary integrity and erases the latter for privacy. We classified transcriber names as positive and interlocutor names as negative, calculating the precision, recall, and F1 scores for each method. Table \ref{tab:precision} demonstrates that \texttt{ProSan}'s F1 score significantly outperforms other methods, indicating its effectiveness in distinguishing names with varying importance to ensure usability.

\begin{table}[ht]
\caption{The precision, recall, and F1 scores of anonymization methods for the name privacy attribute in the summarization task.}
\label{tab:precision}
\centering
\begin{tabular}{c|ccc}
\toprule[1.5pt]
\bf Method & \bf Precision & \bf Recall & \bf $\text{F}_{1}$\\ 
\toprule
HaS & 0.113 & 1 & 0.203\\
\midrule
SanText+ & 0.119 & 0.476 & 0.191 \\
\midrule
\bf $\text{ProSan}^{\diamond}$ & 0.690 & 0.947 & \bf 0.799 \\
\midrule
\bf $\text{ProSan}^{\ast}$ & 0.653 & 0.989 & 0.787 \\
\bottomrule[1.5pt]
\end{tabular}
\end{table}

\textbf{Code Generation}. Table \ref{tab:usability} shows that \texttt{ProSan}'s CodeBLEU scores decreased by only 0.01 on average, while SanText+ dropped by an average of 0.137 due to its poorer readability. HaS decreased by 0.028, which is not significant, but it also failed to effectively protect confidential information such as username, password, and system configuration.

\begin{figure*}[htbp]
    \centering
    \subfloat[Q\&A]{
    \includegraphics[width=0.3\textwidth]{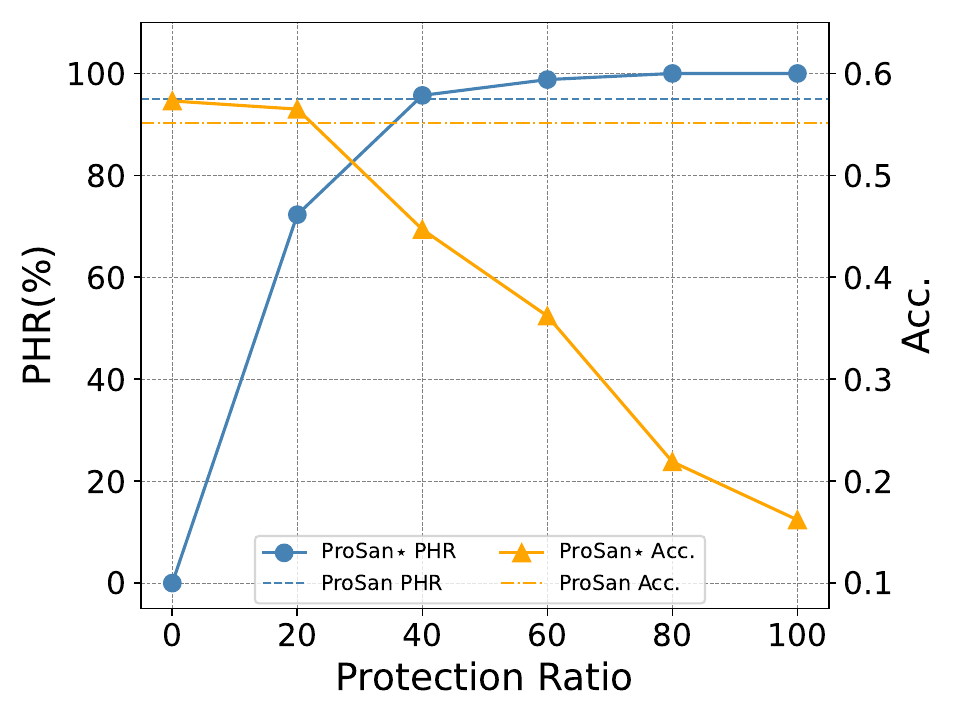}
        \label{fig:sub1}
    }
    \subfloat[Summarization]{
    \includegraphics[width=0.3\textwidth]{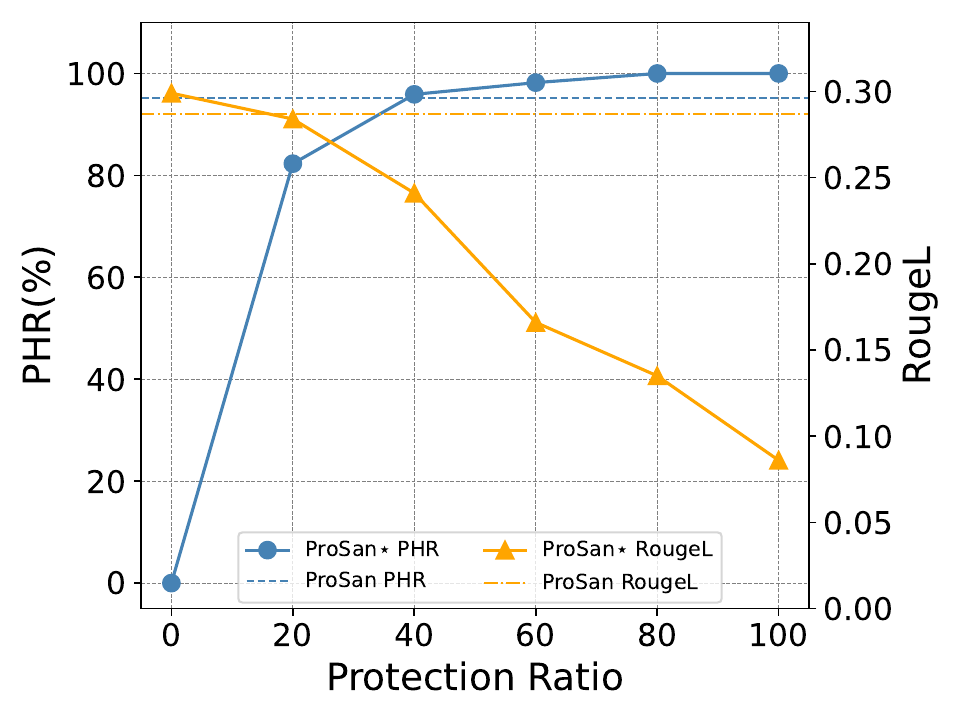}
        \label{fig:sub2}
    }
    \subfloat[Code Generation]{
    \includegraphics[width=0.3\textwidth]{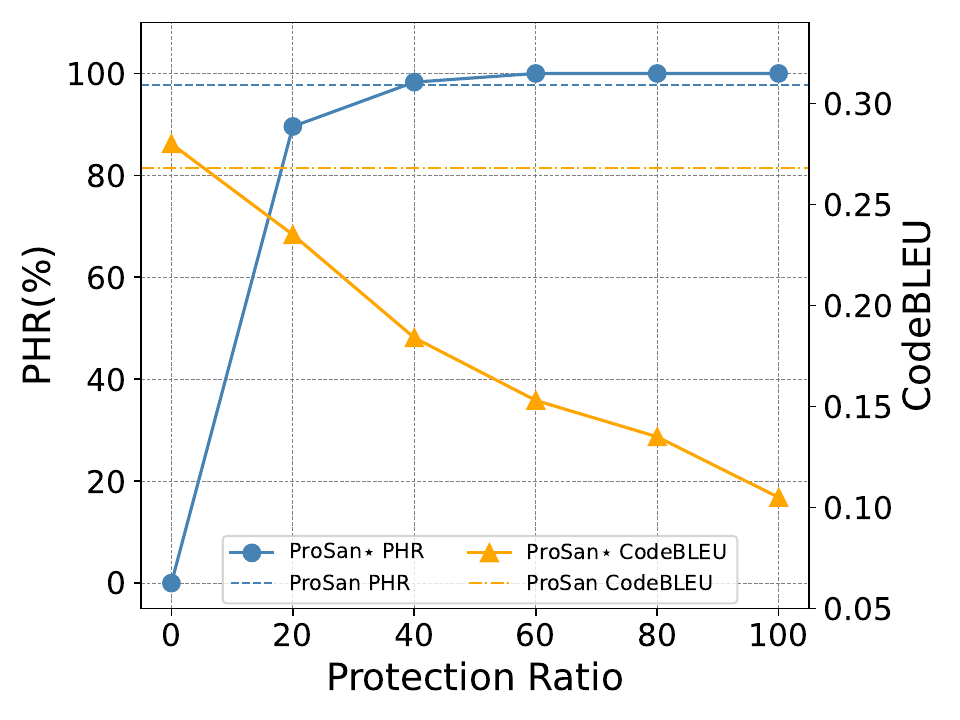}
        \label{fig:sub3}
    }
    \caption{The usability and anonymity performance of different protection ratio calculation strategies across different tasks. $\text{ProSan}^{\star}$ means using a fixed protection ratio.}
    \label{fig:6.3.1_1}
\end{figure*}

\subsubsection{Adaptivity}$\text{ProSan}^{\diamond}$ relies on a billion-parameter LLM, for which we use two RTX 3090 GPU cards, with a runtime memory usage of 30 GB, and the average anonymization time per prompt is 6.52 seconds. In contrast, $\text{ProSan}^{\ast}$ is based on a million-parameter Seq2Seq model, requiring only 2 GB of memory on a laptop equipped with an i9-13905H CPU, reducing anonymization time to 2.03 seconds. $\text{ProSan}^{\diamond}$ serves users with relatively abundant computational resources, offering customizable prompt anonymization, while $\text{ProSan}^{\ast}$ requires fewer resources, offers faster inference, and provides privacy protection with minimal usability trade-off for ordinary users.

\subsection{Ablation Study}
\subsubsection{Protection Ratio}
Figure \ref{fig:6.3.1_1} shows that when we fix the proportion of word protection, it is difficult to achieve a good balance between usability and anonymity due to the varying privacy levels in different prompts. \texttt{ProSan} estimates the overall privacy leakage risk of a prompt in the Anonymized Dataset Generation Module. By dynamically adjusting the protection ratio, it effectively protects privacy while ensuring usability.

\begin{table}[ht]
\caption{The performance of different selection strategies. $\text{ProSan}^{\triangle}$ means proportional random selection of words to be protected.}
\label{tab:performance}
\centering
\begin{tabular}{@{}c|c|cc@{}}
\toprule[1.5pt]
\bf Task & \bf Metric & $\text{ProSan}$ & $\text{ProSan}^{\triangle}$ \\
\toprule
\multirow{2}{*}{Question Answering} & 
PHR(\%) & \bf 95.05 & 51.13 \\
~ & Accuracy & \bf 0.551 & 0.454 \\
\midrule[0.5pt]
\multirow{2}{*}{Text Summarization} & 
PHR(\%) & \bf 95.13 & 45.19 \\
~ & RougeL & \bf 0.287 & 0.226 \\
\midrule[0.5pt]
\multirow{2}{*}{Code Generation} & 
PHR(\%) & \bf 97.68 & 42.76 \\
~ & CodeBLEU & \bf 0.268 & 0.232 \\
\bottomrule[1.5pt]
\end{tabular}
\end{table}

\subsubsection{Selection Strategy}
From Tab. \ref{tab:performance}, we can see that determining which words to protect based on importance is crucial for maintaining both usability and anonymity. When the protection is no longer based on importance and the same proportion of words are randomly replaced, important task-related words may be selected, while unimportant privacy-related words may be overlooked, leading to a decline in both usability and anonymity across all three tasks.

\begin{figure}[t]
\centering
\includegraphics[trim=0mm 6mm 0mm 8mm, clip, width=0.35\textwidth]{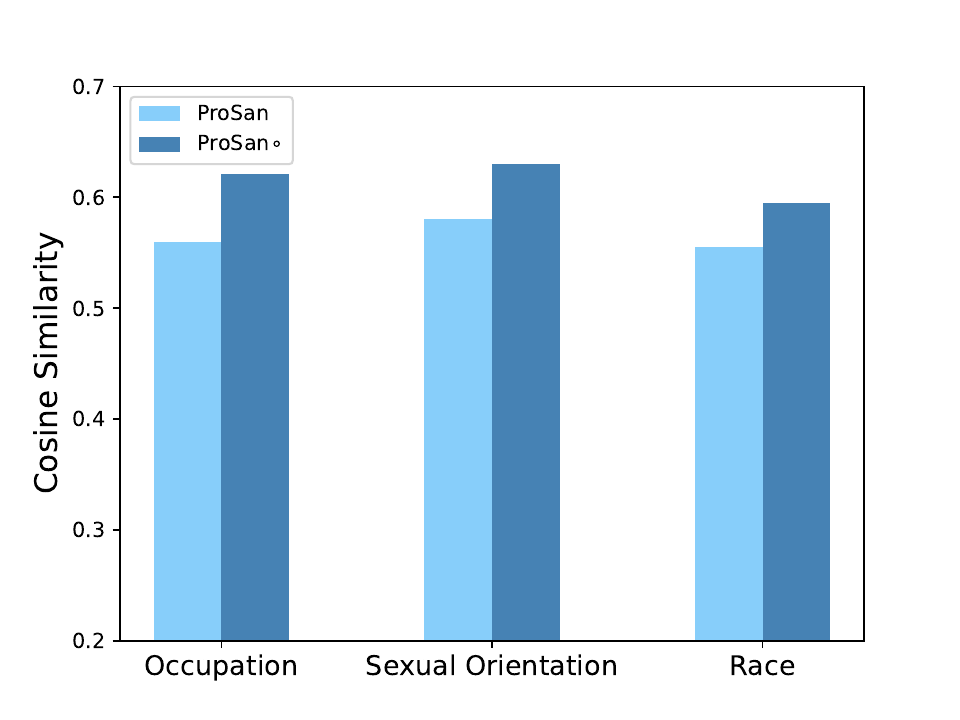}
\caption{The comparison of cosine similarity between anonymized and original prompts under different probability calculation strategies. $\text{ProSan}^\circ$ means directly using the probabilities generated by the masked language model.}
\label{fig:6.3.3_1}
\end{figure}

\subsubsection{Probability Calculation}\label{subsubsec:probability}
For privacy attributes such as occupation, sexual orientation, and race, even if they are successfully captured, the substitute words generated by the masked language model may still risk semantic privacy leakage due to similar meanings. Therefore, we recalculated the sampling probabilities based on the importance, privacy, and similarity of the words to ensure that the substitutes for non-essential terms have minimal similarity to the original words. To verify its effectiveness, we directly used the probabilities generated by the masked language model for sampling substitute words and calculated the cosine similarity between the anonymized and original sentences using Sentence-BERT\cite{reimers2019sentence}. Figure \ref{fig:6.3.3_1} shows that the sentences generated by \texttt{ProSan} using the original sampling probabilities have a higher similarity to the original sentences, indicating a risk of semantic privacy leakage.

\section{Related Work}
\textbf{Preserving privacy in prompt}. Privacy protection for prompts can be divided into the tuning phase and the inference phase. In the tuning phase, the goal is to prevent privacy in the fine-tuning dataset from appearing in prompts. Methods based on sampling and aggregation protect privacy by partitioning local private data into subsets and subsequently aggregating the results\cite{tang2023privacy, hong2023dp}. The PATE\cite{papernot2018scalable} based method trains multiple teacher models using sensitive data and adds Gaussian noise to the predictions to meet privacy requirements\cite{duan2024flocks}. Furthermore, the DP synthetic data generation method employs the DP-SGD\cite{abadi2016deep} framework to fine-tune pre-trained generative language models using private dataset\cite{yue2023synthetic}. In the inference phase, the goal is to prevent the leakage of users' personal private information in prompts. Existing work treats privacy protection as a lexical annotation task, using NER to erase privacy\cite{chen2023hide}. Some work views privacy protection as a text rewriting task, using LLMs to rewrite the original prompt to meet privacy requirements\cite{utpala2023locally}. Other work combines large and small language models, using the large model to complete the main task and the small model to fill in personal privacy information, integrating them to generate the final result.\cite{zhang2024cogenesis}

\textbf{Preserving privacy in text embedding}. This work focuses on protecting privacy in text embeddings rather than in plain text. Some efforts are based on local differential privacy frameworks, which privatize the embeddings of each word\cite{lyu2020towards} or normalize tokens and add noise\cite{plant2021cape}, achieving word-level perturbation. Fine-tuning on sanitized embeddings is also used to improve the performance of the language model\cite{yue2021differential}. Other studies focus on sentence-level perturbation, which protects privacy by cleaning up sentence embeddings\cite{du2023sanitizing} or directly perturbing the embedding matrices during the forward propagation of pre-trained language models\cite{du2023dp}. To protect document embeddings, previous work has selected from the public text embedding space distribution based on the embedding positions of private documents and used an exponential mechanism to generate corresponding privacy embeddings.\cite{meehan2022sentence}

\section{Conclusion}
We propose \texttt{ProSan}, an end-to-end framework designed to protect privacy in user prompts submitted to online LLMs. \texttt{ProSan} dynamically balances usability and anonymity by evaluating the importance and privacy risk of words within the prompts, subsequently anonymizing sensitive information while retaining essential semantic content for task performance. In addition, \texttt{ProSan} has trained a lightweight anonymized model for ordinary users. Compared to the baseline, \texttt{ProSan} effectively minimizes privacy leakage across various tasks without significantly impacting usability. Future work will focus on further exploring other privacy-preserving methods based on self-information and expanding the application scope of \texttt{ProSan} to other privacy-sensitive domains.






%
\bibliographystyle{IEEEtran}
\bibliography{main}

\end{document}

%% file: texs/4.1_2.tex
\redcolorize{11.260188992281613}{Q}
\redcolorize{17.03094568275265}{:}
\redcolorize{9.211570367164395}{ A}
\redcolorize{22.62857967250956}{ 42}
\redcolorize{0.6924908028565245}{-}
\redcolorize{3.635576714996754}{year}
\redcolorize{3.3542523263362907}{-}
\redcolorize{5.424511289042775}{old}
\redcolorize{8.288249296689028}{ man}
\redcolorize{3.96018177883575}{ named}
\redcolorize{1.2551395801774508}{ David}
\redcolorize{5.705835677703239}{ presents}
\redcolorize{4.2847868426747455}{ with}
\redcolorize{12.010387362042847}{ chronic}
\redcolorize{21.64033758926639}{ chest}
\redcolorize{9.247637596479837}{ pain}
\redcolorize{7.07639039169011}{ and}
\redcolorize{24.6303108995167}{ dyspnea}
\redcolorize{12.832720190434971}{,}
\redcolorize{0.025247060520810796}{ which}
\redcolorize{0.660030296472625}{ have}
\redcolorize{3.8303397533001515}{ persisted}
\redcolorize{3.8014859698477963}{ for}
\redcolorize{4.299213734400923}{ 6}
\redcolorize{2.495852268628724}{ months}
\redcolorize{23.479766284354035}{.}
\redcolorize{1.9909110582125082}{ The}
\redcolorize{4.331674240784823}{ symptoms}
\redcolorize{1.6951597778258676}{ typically}
\redcolorize{5.139580177450768}{ intensify}
\redcolorize{0.0}{ during}
\redcolorize{0.19836976123494193}{ physical}
\redcolorize{20.370771117362764}{ exertion}
\redcolorize{22.93154439875929}{.}
\redcolorize{2.586020341917334}{ The}
\redcolorize{4.933996970352737}{ patient}
\redcolorize{2.7266825362475657}{ lives}
\redcolorize{4.587751568924475}{ in}
\redcolorize{1.3921950515761379}{ New}
\redcolorize{4.090023804371349}{ York}
\redcolorize{3.8736204284786844}{ City}
\redcolorize{20.608814830844693}{.}
\redcolorize{7.711173627641925}{ His}
\redcolorize{29.640049051431873}{ familial}
\redcolorize{16.569285147514968}{ history}
\redcolorize{14.131140445790955}{ includes}
\redcolorize{47.38512587463031}{ coronary}
\redcolorize{33.08807617398831}{ artery}
\redcolorize{34.77602250595109}{ disease}
\redcolorize{11.072639399841304}{ in}
\redcolorize{2.1279665296111956}{ his}
\redcolorize{6.030440741542234}{ father}
\redcolorize{15.213157325254272}{,}
\redcolorize{5.107119671066869}{ Steve}
\redcolorize{31.486691192382597}{.}
\redcolorize{18.574623097453653}{ Which}
\redcolorize{12.71730505662555}{ of}
\redcolorize{13.958017745076823}{ the}
\redcolorize{24.042415061674962}{ following}
\redcolorize{38.45487989612638}{ diseases}
\redcolorize{14.232128687874198}{ might}
\redcolorize{17.002091899300297}{ the}
\redcolorize{9.319772055110727}{ patient}
\redcolorize{11.606434393709876}{ have}
\redcolorize{30.851907956430786}{?}
\redcolorize{48.13532424439155}{ \{}
\redcolorize{33.79499386857102}{A}
\redcolorize{62.850753805092694}{:}
\redcolorize{84.63536031162087}{ Coronary}
\redcolorize{51.39580177450769}{ Artery}
\redcolorize{100.0}{ Disease}
\redcolorize{25.311981533578592}{,}
\redcolorize{5.1143331169299575}{ B}
\redcolorize{10.798528457043929}{:}
\redcolorize{37.43778402943086}{ Pleuritis}
\redcolorize{16.80011541513381}{,}
\redcolorize{3.736564957079997}{ C}
\redcolorize{10.582125081151265}{:}
\redcolorize{12.327778980018756}{ Upper}
\redcolorize{52.90341195989324}{ Respiratory}
\redcolorize{17.647695304046742}{ Infection}
\redcolorize{16.19418596263435}{,}
\redcolorize{2.319122844983048}{ D}
\redcolorize{7.675106398326481}{:}
\redcolorize{18.22837769602539}{ Anemia}
\redcolorize{24.864747890067086}{\}}

%% file: texs/4.2_2.tex
\bluecolorize{0.0}{The}
\bluecolorize{38.668082390565374}{ first}
\bluecolorize{100.0}{ astronaut}
\bluecolorize{22.018532321550328}{ to}
\bluecolorize{18.21861108063002}{ land}
\bluecolorize{1.2232545715663408}{ on}
\bluecolorize{0.909669661904847}{ the}
\bluecolorize{2.181019030768143}{ moon}
\bluecolorize{28.275748870252233}{ was}
\bluecolorize{14.97936319596636}{ Neil}
\bluecolorize{6.701715632392297}{ Alden}
\bluecolorize{0.016224861613078515}{ Armstrong}
\bluecolorize{13.07366103933038}{.}

%% file: texs/4.2_3.tex
\bluecolorize{0.0}{The}
\bluecolorize{13.26855337722232}{ name}
\bluecolorize{0.6522557775276417}{ of}
\bluecolorize{10.902036641535032}{ this}
\bluecolorize{56.6679371911769}{ programmer}
\bluecolorize{38.145769474063925}{ working}
\bluecolorize{8.990176183596846}{ in}
\bluecolorize{29.73576609064097}{ Seattle}
\bluecolorize{6.307772671134583}{ is}
\bluecolorize{37.10599298577604}{ Darius}
\bluecolorize{100.0}{ Mae}
\bluecolorize{49.887961176511354}{ Thompson}
\bluecolorize{0.8512960941277484}{.}

%% file: texs/4.2_4.tex
\bluecolorize{0.0}{The}
\bluecolorize{42.352070958851954}{ largest}
\bluecolorize{100.0}{ tropical}
\bluecolorize{41.885256722364794}{ rainforest}
\bluecolorize{6.928687139375766}{ in}
\bluecolorize{19.84854727595564}{ the}
\bluecolorize{7.280056251156919}{ world}
\bluecolorize{13.195510958949763}{ is}
\bluecolorize{36.46626666743314}{ the}
\bluecolorize{0.05987736051081393}{ Amazon}
\bluecolorize{6.9746525794871355}{ rainforest}
\bluecolorize{20.12662957650535}{.}

%% file: texs/4.2_5.tex
\bluecolorize{0.0}{The}
\bluecolorize{37.79663988460973}{ patient}
\bluecolorize{14.76109877062364}{ was}
\bluecolorize{53.94372621326975}{ a}
\bluecolorize{100.0}{ software}
\bluecolorize{13.2438854530219}{ engineer}
\bluecolorize{21.411723077227865}{ who}
\bluecolorize{41.13073411139719}{ worked}
\bluecolorize{17.84591249505078}{ at}
\bluecolorize{72.30503391283588}{ Amazon.com}
\bluecolorize{34.201542151458504}{ Inc.}

%% file: main.bbl
\begin{thebibliography}{10}
\providecommand{\url}[1]{#1}
\csname url@samestyle\endcsname
\providecommand{\newblock}{\relax}
\providecommand{\bibinfo}[2]{#2}
\providecommand{\BIBentrySTDinterwordspacing}{\spaceskip=0pt\relax}
\providecommand{\BIBentryALTinterwordstretchfactor}{4}
\providecommand{\BIBentryALTinterwordspacing}{\spaceskip=\fontdimen2\font plus
\BIBentryALTinterwordstretchfactor\fontdimen3\font minus \fontdimen4\font\relax}
\providecommand{\BIBforeignlanguage}[2]{{%
\expandafter\ifx\csname l@#1\endcsname\relax
\typeout{** WARNING: IEEEtran.bst: No hyphenation pattern has been}%
\typeout{** loaded for the language `#1'. Using the pattern for}%
\typeout{** the default language instead.}%
\else
\language=\csname l@#1\endcsname
\fi
#2}}
\providecommand{\BIBdecl}{\relax}
\BIBdecl

\bibitem{qin2023chatgpt}
C.~Qin, A.~Zhang, Z.~Zhang, J.~Chen, M.~Yasunaga, and D.~Yang, ``Is chatgpt a general-purpose natural language processing task solver?'' in \emph{Proceedings of the 2023 Conference on Empirical Methods in Natural Language Processing}, 2023, pp. 1339--1384.

\bibitem{devlin2019bert}
J.~Devlin, M.-W. Chang, K.~Lee, and K.~Toutanova, ``Bert: Pre-training of deep bidirectional transformers for language understanding,'' in \emph{Proceedings of the 2019 Conference of the North American Chapter of the Association for Computational Linguistics: Human Language Technologies, Volume 1 (Long and Short Papers)}, 2019, pp. 4171--4186.

\bibitem{lester2021power}
B.~Lester, R.~Al-Rfou, and N.~Constant, ``The power of scale for parameter-efficient prompt tuning,'' in \emph{Proceedings of the 2021 Conference on Empirical Methods in Natural Language Processing}, 2021, pp. 3045--3059.

\bibitem{radford2019language}
A.~Radford, J.~Wu, R.~Child, D.~Luan, D.~Amodei, I.~Sutskever \emph{et~al.}, ``Language models are unsupervised multitask learners,'' \emph{OpenAI blog}, vol.~1, no.~8, p.~9, 2019.

\bibitem{touvron2023llama}
H.~Touvron, T.~Lavril, G.~Izacard, X.~Martinet, M.-A. Lachaux, T.~Lacroix, B.~Rozi{\`e}re, N.~Goyal, E.~Hambro, F.~Azhar \emph{et~al.}, ``Llama: Open and efficient foundation language models,'' \emph{arXiv preprint arXiv:2302.13971}, 2023.

\bibitem{edemacu2024privacy}
K.~Edemacu and X.~Wu, ``Privacy preserving prompt engineering: A survey,'' \emph{arXiv preprint arXiv:2404.06001}, 2024.

\bibitem{openai2024privacy}
\url{https://openai.com/policies/privacy-policy}, 2024, [Accessed 06-05-2024].

\bibitem{li2023multi}
H.~Li, D.~Guo, W.~Fan, M.~Xu, J.~Huang, F.~Meng, and Y.~Song, ``Multi-step jailbreaking privacy attacks on chatgpt,'' in \emph{Findings of the Association for Computational Linguistics: EMNLP 2023}, 2023, pp. 4138--4153.

\bibitem{carlini2021extracting}
N.~Carlini, F.~Tramer, E.~Wallace, M.~Jagielski, A.~Herbert-Voss, K.~Lee, A.~Roberts, T.~Brown, D.~Song, U.~Erlingsson \emph{et~al.}, ``Extracting training data from large language models,'' in \emph{30th USENIX Security Symposium (USENIX Security 21)}, 2021, pp. 2633--2650.

\bibitem{gupta2022recovering}
S.~Gupta, Y.~Huang, Z.~Zhong, T.~Gao, K.~Li, and D.~Chen, ``Recovering private text in federated learning of language models,'' \emph{Advances in neural information processing systems}, vol.~35, pp. 8130--8143, 2022.

\bibitem{chu2024conversation}
J.~Chu, Z.~Sha, M.~Backes, and Y.~Zhang, ``Conversation reconstruction attack against gpt models,'' \emph{arXiv preprint arXiv:2402.02987}, 2024.

\bibitem{zhang2024cogenesis}
K.~Zhang, J.~Wang, E.~Hua, B.~Qi, N.~Ding, and B.~Zhou, ``Cogenesis: A framework collaborating large and small language models for secure context-aware instruction following,'' \emph{arXiv preprint arXiv:2403.03129}, 2024.

\bibitem{kan2023protecting}
Z.~Kan, L.~Qiao, H.~Yu, L.~Peng, Y.~Gao, and D.~Li, ``Protecting user privacy in remote conversational systems: A privacy-preserving framework based on text sanitization,'' \emph{arXiv preprint arXiv:2306.08223}, 2023.

\bibitem{xu2023survey}
C.~Xu and J.~McAuley, ``A survey on model compression and acceleration for pretrained language models,'' in \emph{Proceedings of the AAAI Conference on Artificial Intelligence}, vol.~37, no.~9, 2023, pp. 10\,566--10\,575.

\bibitem{huang2020texthide}
Y.~Huang, Z.~Song, D.~Chen, K.~Li, and S.~Arora, ``Texthide: Tackling data privacy in language understanding tasks,'' in \emph{Findings of the Association for Computational Linguistics: EMNLP 2020}, 2020, pp. 1368--1382.

\bibitem{zhou2022textfusion}
X.~Zhou, J.~Lu, T.~Gui, R.~Ma, Z.~Fei, Y.~Wang, Y.~Ding, Y.~Cheung, Q.~Zhang, and X.-J. Huang, ``Textfusion: Privacy-preserving pre-trained model inference via token fusion,'' in \emph{Proceedings of the 2022 Conference on Empirical Methods in Natural Language Processing}, 2022, pp. 8360--8371.

\bibitem{zhou2023textmixer}
X.~Zhou, Y.~Lu, R.~Ma, T.~Gui, Q.~Zhang, and X.-J. Huang, ``Textmixer: Mixing multiple inputs for privacy-preserving inference,'' in \emph{Findings of the Association for Computational Linguistics: EMNLP 2023}, 2023, pp. 3749--3762.

\bibitem{pan2020privacy}
X.~Pan, M.~Zhang, S.~Ji, and M.~Yang, ``Privacy risks of general-purpose language models,'' in \emph{2020 IEEE Symposium on Security and Privacy (SP)}.\hskip 1em plus 0.5em minus 0.4em\relax IEEE, 2020, pp. 1314--1331.

\bibitem{zhang2024privacyasst}
X.~Zhang, H.~Xu, Z.~Ba, Z.~Wang, Y.~Hong, J.~Liu, Z.~Qin, and K.~Ren, ``Privacyasst: Safeguarding user privacy in tool-using large language model agents,'' \emph{IEEE Transactions on Dependable and Secure Computing}, 2024.

\bibitem{dwork2006differential}
C.~Dwork, ``Differential privacy,'' in \emph{International colloquium on automata, languages, and programming}.\hskip 1em plus 0.5em minus 0.4em\relax Springer, 2006, pp. 1--12.

\bibitem{yue2021differential}
X.~Yue, M.~Du, T.~Wang, Y.~Li, H.~Sun, and S.~S. Chow, ``Differential privacy for text analytics via natural text sanitization,'' \emph{arXiv preprint arXiv:2106.01221}, 2021.

\bibitem{du2023sanitizing}
M.~Du, X.~Yue, S.~S. Chow, and H.~Sun, ``Sanitizing sentence embeddings (and labels) for local differential privacy,'' in \emph{Proceedings of the ACM Web Conference 2023}, 2023, pp. 2349--2359.

\bibitem{meehan2022sentence}
C.~Meehan, K.~Mrini, and K.~Chaudhuri, ``Sentence-level privacy for document embeddings,'' in \emph{Proceedings of the 60th Annual Meeting of the Association for Computational Linguistics (Volume 1: Long Papers)}, 2022, pp. 3367--3380.

\bibitem{feyisetan2020privacy}
O.~Feyisetan, B.~Balle, T.~Drake, and T.~Diethe, ``Privacy-and utility-preserving textual analysis via calibrated multivariate perturbations,'' in \emph{Proceedings of the 13th international conference on web search and data mining}, 2020, pp. 178--186.

\bibitem{krishna2021adept}
S.~Krishna, R.~Gupta, and C.~Dupuy, ``Adept: Auto-encoder based differentially private text transformation,'' in \emph{Proceedings of the 16th Conference of the European Chapter of the Association for Computational Linguistics: Main Volume}, 2021, pp. 2435--2439.

\bibitem{mattern2022limits}
J.~Mattern, B.~Weggenmann, and F.~Kerschbaum, ``The limits of word level differential privacy,'' in \emph{Findings of the Association for Computational Linguistics: NAACL 2022}, 2022, pp. 867--881.

\bibitem{lison2021anonymisation}
P.~Lison, I.~Pil{\'a}n, D.~S{\'a}nchez, M.~Batet, and L.~{\O}vrelid, ``Anonymisation models for text data: State of the art, challenges and future directions,'' in \emph{Proceedings of the 59th Annual Meeting of the Association for Computational Linguistics and the 11th International Joint Conference on Natural Language Processing (Volume 1: Long Papers)}, 2021, pp. 4188--4203.

\bibitem{nadeau2007survey}
D.~Nadeau and S.~Sekine, ``A survey of named entity recognition and classification,'' \emph{Lingvisticae Investigationes}, vol.~30, no.~1, pp. 3--26, 2007.

\bibitem{chen2023hide}
Y.~Chen, T.~Li, H.~Liu, and Y.~Yu, ``Hide and seek (has): A lightweight framework for prompt privacy protection,'' \emph{arXiv preprint arXiv:2309.03057}, 2023.

\bibitem{utpala2023locally}
S.~Utpala, S.~Hooker, and P.-Y. Chen, ``Locally differentially private document generation using zero shot prompting,'' in \emph{Findings of the Association for Computational Linguistics: EMNLP 2023}, 2023, pp. 8442--8457.

\bibitem{zou2023universal}
A.~Zou, Z.~Wang, J.~Z. Kolter, and M.~Fredrikson, ``Universal and transferable adversarial attacks on aligned language models,'' \emph{arXiv preprint arXiv:2307.15043}, 2023.

\bibitem{vaswani2017attention}
A.~Vaswani, N.~Shazeer, N.~Parmar, J.~Uszkoreit, L.~Jones, A.~N. Gomez, {\L}.~Kaiser, and I.~Polosukhin, ``Attention is all you need,'' \emph{Advances in neural information processing systems}, vol.~30, 2017.

\bibitem{liu2019roberta}
Y.~Liu, M.~Ott, N.~Goyal, J.~Du, M.~Joshi, D.~Chen, O.~Levy, M.~Lewis, L.~Zettlemoyer, and V.~Stoyanov, ``Roberta: A robustly optimized bert pretraining approach,'' \emph{arXiv preprint arXiv:1907.11692}, 2019.

\bibitem{brown2020language}
T.~Brown, B.~Mann, N.~Ryder, M.~Subbiah, J.~D. Kaplan, P.~Dhariwal, A.~Neelakantan, P.~Shyam, G.~Sastry, A.~Askell \emph{et~al.}, ``Language models are few-shot learners,'' \emph{Advances in neural information processing systems}, vol.~33, pp. 1877--1901, 2020.

\bibitem{lewis2020bart}
M.~Lewis, Y.~Liu, N.~Goyal, M.~Ghazvininejad, A.~Mohamed, O.~Levy, V.~Stoyanov, and L.~Zettlemoyer, ``Bart: Denoising sequence-to-sequence pre-training for natural language generation, translation, and comprehension,'' in \emph{Proceedings of the 58th Annual Meeting of the Association for Computational Linguistics}, 2020, pp. 7871--7880.

\bibitem{chung2024scaling}
H.~W. Chung, L.~Hou, S.~Longpre, B.~Zoph, Y.~Tay, W.~Fedus, Y.~Li, X.~Wang, M.~Dehghani, S.~Brahma \emph{et~al.}, ``Scaling instruction-finetuned language models,'' \emph{Journal of Machine Learning Research}, vol.~25, no.~70, pp. 1--53, 2024.

\bibitem{wei2022emergent}
J.~Wei, Y.~Tay, R.~Bommasani, C.~Raffel, B.~Zoph, S.~Borgeaud, D.~Yogatama, M.~Bosma, D.~Zhou, D.~Metzler \emph{et~al.}, ``Emergent abilities of large language models,'' \emph{arXiv preprint arXiv:2206.07682}, 2022.

\bibitem{ouyang2022training}
L.~Ouyang, J.~Wu, X.~Jiang, D.~Almeida, C.~Wainwright, P.~Mishkin, C.~Zhang, S.~Agarwal, K.~Slama, A.~Ray \emph{et~al.}, ``Training language models to follow instructions with human feedback,'' \emph{Advances in neural information processing systems}, vol.~35, pp. 27\,730--27\,744, 2022.

\bibitem{tirumala2022memorization}
K.~Tirumala, A.~Markosyan, L.~Zettlemoyer, and A.~Aghajanyan, ``Memorization without overfitting: Analyzing the training dynamics of large language models,'' \emph{Advances in Neural Information Processing Systems}, vol.~35, pp. 38\,274--38\,290, 2022.

\bibitem{carlini2022membership}
N.~Carlini, S.~Chien, M.~Nasr, S.~Song, A.~Terzis, and F.~Tramer, ``Membership inference attacks from first principles,'' in \emph{2022 IEEE Symposium on Security and Privacy (SP)}.\hskip 1em plus 0.5em minus 0.4em\relax IEEE, 2022, pp. 1897--1914.

\bibitem{duan2024flocks}
H.~Duan, A.~Dziedzic, N.~Papernot, and F.~Boenisch, ``Flocks of stochastic parrots: Differentially private prompt learning for large language models,'' \emph{Advances in Neural Information Processing Systems}, vol.~36, 2024.

\bibitem{hong2023dp}
J.~Hong, J.~T. Wang, C.~Zhang, L.~Zhangheng, B.~Li, and Z.~Wang, ``Dp-opt: Make large language model your differentially-private prompt engineer,'' in \emph{The Twelfth International Conference on Learning Representations}, 2023.

\bibitem{lin2024promptcrypt}
G.~Lin, W.~Hua, and Y.~Zhang, ``Promptcrypt: Prompt encryption for secure communication with large language models,'' \emph{arXiv preprint arXiv:2402.05868}, 2024.

\bibitem{dou2023reducing}
Y.~Dou, I.~Krsek, T.~Naous, A.~Kabra, S.~Das, A.~Ritter, and W.~Xu, ``Reducing privacy risks in online self-disclosures with language models,'' \emph{arXiv preprint arXiv:2311.09538}, 2023.

\bibitem{bevendorff2019heuristic}
J.~Bevendorff, M.~Potthast, M.~Hagen, and B.~Stein, ``Heuristic authorship obfuscation,'' in \emph{Proceedings of the 57th Annual Meeting of the Association for Computational Linguistics}, 2019, pp. 1098--1108.

\bibitem{mahmood2019girl}
A.~Mahmood, F.~Ahmad, Z.~Shafiq, P.~Srinivasan, and F.~Zaffar, ``A girl has no name: Automated authorship obfuscation using mutant-x,'' \emph{Proceedings on Privacy Enhancing Technologies}, 2019.

\bibitem{shetty2018a4nt}
R.~Shetty, B.~Schiele, and M.~Fritz, ``$\{$A4NT$\}$: Author attribute anonymity by adversarial training of neural machine translation,'' in \emph{27th USENIX Security Symposium (USENIX Security 18)}, 2018, pp. 1633--1650.

\bibitem{rosado2023pii}
E.~J. Rosado, ``Pii-codex: a python library for pii detection, categorization, and severity assessment,'' \emph{The Journal of Open Source Software}, vol.~8, no.~86, p. 5402, 2023.

\bibitem{herwanto2021named}
G.~B. Herwanto, G.~Quirchmayr, and A.~M. Tjoa, ``A named entity recognition based approach for privacy requirements engineering,'' in \emph{2021 IEEE 29th International Requirements Engineering Conference Workshops (REW)}.\hskip 1em plus 0.5em minus 0.4em\relax IEEE, 2021, pp. 406--411.

\bibitem{cumby2011machine}
C.~Cumby and R.~Ghani, ``A machine learning based system for semi-automatically redacting documents,'' in \emph{Proceedings of the AAAI Conference on Artificial Intelligence}, vol.~25, no.~2, 2011, pp. 1628--1635.

\bibitem{feyisetan2019leveraging}
O.~Feyisetan, T.~Diethe, and T.~Drake, ``Leveraging hierarchical representations for preserving privacy and utility in text,'' in \emph{2019 IEEE International Conference on Data Mining (ICDM)}.\hskip 1em plus 0.5em minus 0.4em\relax IEEE, 2019, pp. 210--219.

\bibitem{qu2021natural}
C.~Qu, W.~Kong, L.~Yang, M.~Zhang, M.~Bendersky, and M.~Najork, ``Natural language understanding with privacy-preserving bert,'' in \emph{Proceedings of the 30th ACM International Conference on Information \& Knowledge Management}, 2021, pp. 1488--1497.

\bibitem{nissenbaum2004privacy}
H.~Nissenbaum, ``Privacy as contextual integrity,'' \emph{Wash. L. Rev.}, vol.~79, p. 119, 2004.

\bibitem{jin2020disease}
D.~Jin, E.~Pan, N.~Oufattole, W.-H. Weng, H.~Fang, and P.~Szolovits, ``What disease does this patient have? a large-scale open domain question answering dataset from medical exams,'' \emph{arXiv preprint arXiv:2009.13081}, 2020.

\bibitem{europaRegulation2016679}
``{R}egulation - 2016/679 - {E}{N} - gdpr - {E}{U}{R}-{L}ex --- data.europa.eu,'' \url{http://data.europa.eu/eli/reg/2016/679/oj}, [Accessed 10-05-2024].

\bibitem{shannon2001mathematical}
C.~E. Shannon, ``A mathematical theory of communication,'' \emph{ACM SIGMOBILE mobile computing and communications review}, vol.~5, no.~1, pp. 3--55, 2001.

\bibitem{li2023compressing}
Y.~Li, B.~Dong, F.~Guerin, and C.~Lin, ``Compressing context to enhance inference efficiency of large language models,'' in \emph{Proceedings of the 2023 Conference on Empirical Methods in Natural Language Processing}, 2023, pp. 6342--6353.

\bibitem{bunescu2022distribution}
R.~Bunescu and O.~O. Uduehi, ``Distribution-based measures of surprise for creative language: Experiments with humor and metaphor,'' in \emph{Proceedings of the 3rd Workshop on Figurative Language Processing (FLP)}, 2022, pp. 68--78.

\bibitem{ai4privacy_2023}
\BIBentryALTinterwordspacing
{ai4Privacy}, ``pii-masking-43k (revision c47c98d),'' 2023. [Online]. Available: \url{https://huggingface.co/datasets/ai4privacy/pii-masking-43k}
\BIBentrySTDinterwordspacing

\bibitem{mikolov2013efficient}
T.~Mikolov, K.~Chen, G.~Corrado, and J.~Dean, ``Efficient estimation of word representations in vector space,'' \emph{arXiv preprint arXiv:1301.3781}, 2013.

\bibitem{pennington2014glove}
J.~Pennington, R.~Socher, and C.~D. Manning, ``Glove: Global vectors for word representation,'' in \emph{Proceedings of the 2014 conference on empirical methods in natural language processing (EMNLP)}, 2014, pp. 1532--1543.

\bibitem{peters2018deep}
M.~E. Peters, M.~Neumann, M.~Iyyer, M.~Gardner, C.~Clark, K.~Lee, and L.~Zettlemoyer, ``Deep contextualized word representations,'' in \emph{Proceedings of the 2018 Conference of the North American Chapter of the Association for Computational Linguistics: Human Language Technologies, Volume 1 (Long Papers)}, 2018, pp. 2227--2237.

\bibitem{miller1995wordnet}
G.~A. Miller, ``Wordnet: a lexical database for english,'' \emph{Communications of the ACM}, vol.~38, no.~11, pp. 39--41, 1995.

\bibitem{ge2022edgeformer}
T.~Ge, S.-Q. Chen, and F.~Wei, ``Edgeformer: A parameter-efficient transformer for on-device seq2seq generation,'' in \emph{Proceedings of the 2022 Conference on Empirical Methods in Natural Language Processing}, 2022, pp. 10\,786--10\,798.

\bibitem{jin2021disease}
D.~Jin, E.~Pan, N.~Oufattole, W.-H. Weng, H.~Fang, and P.~Szolovits, ``What disease does this patient have? a large-scale open domain question answering dataset from medical exams,'' \emph{Applied Sciences}, vol.~11, no.~14, p. 6421, 2021.

\bibitem{gliwa2019samsum}
B.~Gliwa, I.~Mochol, M.~Biesek, and A.~Wawer, ``Samsum corpus: A human-annotated dialogue dataset for abstractive summarization,'' \emph{arXiv preprint arXiv:1911.12237}, 2019.

\bibitem{iamtarun_2023}
\BIBentryALTinterwordspacing
{iamtarun}, ``python-code-instructions-18k-alpaca,'' 2023. [Online]. Available: \url{https://huggingface.co/datasets/iamtarun/python_code_instructions_18k_alpaca}
\BIBentrySTDinterwordspacing

\bibitem{wang2023self}
Y.~Wang, Y.~Kordi, S.~Mishra, A.~Liu, N.~A. Smith, D.~Khashabi, and H.~Hajishirzi, ``Self-instruct: Aligning language models with self-generated instructions,'' in \emph{Proceedings of the 61st Annual Meeting of the Association for Computational Linguistics (Volume 1: Long Papers)}, 2023, pp. 13\,484--13\,508.

\bibitem{reimers2019sentence}
N.~Reimers and I.~Gurevych, ``Sentence-bert: Sentence embeddings using siamese bert-networks,'' in \emph{Proceedings of the 2019 Conference on Empirical Methods in Natural Language Processing and the 9th International Joint Conference on Natural Language Processing (EMNLP-IJCNLP)}, 2019, pp. 3982--3992.

\bibitem{tang2023privacy}
X.~Tang, R.~Shin, H.~A. Inan, A.~Manoel, F.~Mireshghallah, Z.~Lin, S.~Gopi, J.~Kulkarni, and R.~Sim, ``Privacy-preserving in-context learning with differentially private few-shot generation,'' \emph{arXiv preprint arXiv:2309.11765}, 2023.

\bibitem{papernot2018scalable}
N.~Papernot, S.~Song, I.~Mironov, A.~Raghunathan, K.~Talwar, and {\'U}.~Erlingsson, ``Scalable private learning with pate,'' \emph{arXiv preprint arXiv:1802.08908}, 2018.

\bibitem{abadi2016deep}
M.~Abadi, A.~Chu, I.~Goodfellow, H.~B. McMahan, I.~Mironov, K.~Talwar, and L.~Zhang, ``Deep learning with differential privacy,'' in \emph{Proceedings of the 2016 ACM SIGSAC conference on computer and communications security}, 2016, pp. 308--318.

\bibitem{yue2023synthetic}
X.~Yue, H.~Inan, X.~Li, G.~Kumar, J.~McAnallen, H.~Shajari, H.~Sun, D.~Levitan, and R.~Sim, ``Synthetic text generation with differential privacy: A simple and practical recipe,'' in \emph{Proceedings of the 61st Annual Meeting of the Association for Computational Linguistics (Volume 1: Long Papers)}, 2023, pp. 1321--1342.

\bibitem{lyu2020towards}
L.~Lyu, Y.~Li, X.~He, and T.~Xiao, ``Towards differentially private text representations,'' in \emph{Proceedings of the 43rd International ACM SIGIR Conference on Research and Development in Information Retrieval}, 2020, pp. 1813--1816.

\bibitem{plant2021cape}
R.~Plant, D.~Gkatzia, and V.~Giuffrida, ``Cape: Context-aware private embeddings for private language learning,'' in \emph{Proceedings of the 2021 Conference on Empirical Methods in Natural Language Processing}, 2021, pp. 7970--7978.

\bibitem{du2023dp}
M.~Du, X.~Yue, S.~S. Chow, T.~Wang, C.~Huang, and H.~Sun, ``Dp-forward: Fine-tuning and inference on language models with differential privacy in forward pass,'' in \emph{Proceedings of the 2023 ACM SIGSAC Conference on Computer and Communications Security}, 2023, pp. 2665--2679.

\end{thebibliography}
